\documentclass[twoside,journal]{IEEEtran}

\usepackage{cite,graphicx,amssymb,amsmath,color,textcomp,tabularx}
\newtheorem{theorem}{\underline{Theorem}}%[section]
\newtheorem{lemma}{\underline{Lemma}}%[section]
\newtheorem{remark}{\underline{Remark}}%[section]

\IEEEoverridecommandlockouts
\newcounter{mytempeqncnt}

\usepackage{graphicx}
\usepackage{epstopdf}
\usepackage{stfloats}
\usepackage{algorithm,algorithmic}

\usepackage{bm}
\usepackage{multirow}
\usepackage[caption=false,font=footnotesize]{subfig}

\usepackage{enumerate,extarrows}

\def \Tr {\mathrm{Tr}}
\def \diag {\mathrm{diag}}
\def \st {\mathrm{s.t.}}
\def \L {\mathbf{\Lambda}_z}

\newcommand{\tabincell}[2]{\begin{tabular}{@{}#1@{}}#2\end{tabular}}

\begin{document}

\title{Physical Layer Security in MIMO Backscatter Wireless Systems}
\author{Qian~Yang,~\IEEEmembership{Student Member,~IEEE,}
	Hui-Ming~Wang,~\IEEEmembership{Senior Member,~IEEE,}\\
	Yi~Zhang,~\IEEEmembership{Student Member,~IEEE,}
	and~Zhu~Han,~\IEEEmembership{Fellow,~IEEE}
	\thanks{The work of Q. Yang, H.-M. Wang, and Y. Zhang was partially supported by the National Natural Science Foundation of China under Grant 61671364, the Foundation for the Author of National Excellent Doctoral Dissertation of China under Grant 201340,  and the Young Talent Support Fund of Science and Technology of Shaanxi Province under Grant 2015KJXX-01.
	The work of Z. Han was supported in part by the U.S. NSF ECCS-1547201, CCF-1456921, CNS-1443917, ECCS-1405121, and NSFC 61428101. \emph{(Corresponding author: Hui-Ming Wang.)}
		}
	\thanks{Q. Yang, H.-M. Wang, and Y. Zhang are with the School of Electronic and Information Engineering, Xi'an Jiaotong University, Xi'an 710049, China, and also with the MOE Key Lab for Intelligent Networks and Network Security, Xi'an Jiaotong University, Xi'an 710049, China (e-mail: yangq36@gmail.com; xjbswhm@gmail.com; yi.zhang.cn@outlook.com).}
	\thanks{Z. Han is with the Department of Electrical and Computer Engineering, University of Houston, Houston, TX 77004 USA (e-mail: zhan2@uh.edu).}
	}

\maketitle

\begin{abstract}
	Backscatter wireless communication is an emerging technique widely used in low-cost and low-power wireless systems, especially in passive radio frequency identification (RFID) systems. 
	Recently, the requirement of high data rates, data reliability, and security drives the development of RFID systems, which motivates our investigation on the physical layer security of a multiple-input multiple-output (MIMO) RFID system.
	In this paper, we propose a noise-injection precoding strategy to safeguard the system security with the resource-constrained nature of the backscatter system taken into consideration.
	We first consider a multi-antenna RFID tag case and investigate the secrecy rate maximization (SRM) problem by jointly optimizing the energy supply power and the precoding matrix of the injected artificial noise at the RFID reader. 
	We exploit the alternating optimization method and the sequential parametric convex approximation method, respectively, to tackle the non-convex SRM problem and show an interesting fact that the two methods are actually equivalent for our SRM problem with the convergence of a Karush-Kuhn-Tucker (KKT) point. 
	To facilitate the practical implementation for resource-constrained RFID devices, we propose a fast algorithm based on projected gradient. 
	We also consider a single-antenna RFID tag case and develop a low-complexity algorithm which yields the global optimal solution.
	Simulation results show the superiority of our proposed algorithms in terms of the secrecy rate and computational complexity.
\end{abstract}

\begin{IEEEkeywords}MIMO backscatter wireless communication, RFID, physical layer security, artificial noise, optimization.
\end{IEEEkeywords}

%\newpage

\section{Introduction}
%Part1.	Usage of backscatter wireless communications in RFID systems, and the bright future of RFID systems
\IEEEPARstart{B}{ackscatter} wireless communication, remarkable for its low energy consumption and low product cost, is an emerging technology which is widely used\cite{Landt2005,Want2006,Dobkin2007,Boyer2014}. One of its most prominent applications is in radio frequency identification (RFID) systems. RFID enables identification from a distance, thereby facilitating the handling of manufactured goods and materials\cite{Landt2005,Want2006}. 
By employing backscatter modulation\cite{Dobkin2007} to send back the data and on-tag power harvesting\cite{Boyer2014} to supply the power, RFID is promoted by its longevity, efficacy, and energy efficiency.
It is believed that RFID will become one of the most crucial techniques to realize the Internet of Things (IoT)\cite{Kellogg2014}, which allows objects to be sensed and creates more efficient interactions between the physical world and computer-based systems.

%Part3
As a contactless technology in a short range, RFID systems are expected to fulfill the aim of reducing handling time despite the augment of data stored in RFID tags\cite{Pillin2008,Gossar2011}. 
Concerning this higher expectation of the data rate and data reliability for novel RFID applications, the implementation of the multiple-input multiple-output (MIMO) scheme appears to be effective and promising\cite{Karmakar2010,Zheng2016}, which attracts considerable research interests\cite{Ingram2001,Griffin2008,Zheng2012,Boyer2013,He2015}. 
%As seen from the wireless communications in Wi-Fi and cellular networks, multiple antenna techniques are efficient in improving the data rate and data reliability\cite{Tse2005}. 
It is shown in \cite{Ingram2001} that adopting multiple antennas can extend the coverage of backscatter RFID systems and improve system capacity under the spatial multiplexing configuration.
%Furthermore, adopting multiple antennas at the tag was investigated in \cite{Griffin2008}, and it was shown that pinhole diversity gains are available to RF tags with more than one antenna. 
%In \cite{Zheng2012}, the authors studied the reader's transmit signal design problem and proposed two spatial matching transmission schemes that can greatly improve the performance of MIMO RFID systems. 
Furthermore, 
%equipping with multiple antennas at the RFID tag or reader was investigated in \cite{Griffin2008,Zheng2012}, and 
the authors in \cite{Griffin2008,Zheng2012} show that multi-antenna techniques can significantly improve the data reliability of the RFID system. 
The space-time coding scheme is explored in \cite{Boyer2013,He2015} for MIMO RFID backscatter systems. %, and it was shown that adopting multiple antennas at both the reader and the tag can significantly improve the reliability of the system. 
%The MIMO scheme has been extensively investigated and recognized as a promising technique in the backscatter wireless communication system.
Apart from the above analytical studies, several real experiments concerning multi-antenna RFID tags have also been conducted\cite{Akbar2012,Denicke2012,Trotter2012}. The measurement results in \cite{Akbar2012} show that read range can be improved when multiple antennas, instead of a single antenna, are equipped at the RFID tag. In addition, the authors in \cite{Denicke2012} propose a method for channel measurements in MIMO RFID systems. The authors in \cite{Trotter2012} showcase two multi-antenna techniques for RFID tags operating at 5.8 GHz. The MIMO scheme has been extensively investigated and recognized as an efficient approach to further extending the information-carrying ability of RFID \cite{Zheng2016}.

%In addition, security concerns including content privacy, data protection, etc. also make a great challenge to the design of RFID systems\cite{Chien2007}.
Due to the widespread deployment of RFID tags, the privacy concern for users, such as clandestine physical tracking and personal information protecting, also makes a great challenge to the design of RFID systems, because the transmission is vulnerable to eavesdropping due to the broadcast nature of backscatter communication \cite{Garfinkel2005,Juels2006,Chien2007}. 
%Meanwhile, due to the broadcast nature of backscatter systems, the secure transmission against malicious attacks, such as eavesdropping, also becomes a great challenge. 
Most of previous works concerning RFID security issues mainly focus on lightweight cryptography such as in\cite{Chien2007,Eisenbarth2007,Vahedi2011}. 
%Although these approaches, to some extent, can guarantee the security of RFID systems against malicious attacks
However, they still have some restrictions on the secret key generation and distribution from eavesdropping and practical limitations in terms of size, cost, and computation\cite{Vahedi2011,Saad2014}. 
Fortunately, %besides the field of cryptography
in recent years physical layer security (PLS), as an alternative or complement to cryptography, has drawn considerable attention in strengthening the security of wireless communications since perfect secrecy is provided.
The theoretical basis for PLS approaches lies in the notion of the \emph{secrecy capacity/rate}, which was pioneered by Wyner in \cite{Wyner1975}.
%, where the author introduced the wiretap channel and defined the concept of secrecy capacity as a metric to evaluate the secure performance of a system against eavesdropping. 
Since then, wealth of relevant research has achieved a significant success in the security of conventional wireless communication systems\cite{Oggier2011,Bloch2011,Mukherjee2014,Negi2005,Li2013,Li2013a,Wang2015b,Wang2016}. 
The main idea of PLS approaches, in addition to exploiting the randomness inherent to wireless channels\cite{Mukherjee2014}, is to manually construct \emph{equivalent channels} via signal design and power allocation such that the superiority of the equivalent legitimate channel to the equivalent wiretap channel can be established\cite{Wang2015}.
%The main idea of realizing a positive secrecy capacity is to exploit the randomness inherent to wireless channels or to manually construct equivalent legitimate and wiretap channels via transmission signal design and power allocation such that the destination has a better equivalent channel than the eavesdropper\cite{Mukherjee2014}.
One of promising approaches is to send an artificially generated noise to deteriorate the channel condition of eavesdroppers. This conception of applying artificial noise (AN) to enhance the secure transmission is first introduced in \cite{Negi2005}, and it is further studied in \cite{Li2013,Li2013a,Wang2015b,Wang2016}. These studies have also been generalized to the cooperative relay system\cite{Li2011,Wang2013,Wang2015c,Wang2015}. 
%Isotropic AN and spatially selective AN can be implemented, respectively, according to whether the knowledge of channel state information (CSI) is available.
%These studies on the security of conventional wireless systems are potential to be used as reference in strengthening the security of backscatter systems. 

However, there is only little work to study the security of  backscatter systems from the perspective of PLS. 
%Recently, the PLS technique was exploited 
In \cite{Saad2014}, a physical layer noise injection scheme is proposed to strengthen the security of backscatter wireless systems under a single-input single-output (SISO) system setting where all terminals employ a single antenna, and it is shown that the proposed approach yields significant performance gains. 
To the best of our knowledge, no work has been done on the PLS of a MIMO backscatter system, even though the MIMO technique is promising in enhancing the security due to extra spatial degrees of freedom provided by multiple antennas\cite{Li2013}.
Note that the PLS approaches for the conventional MIMO system cannot be directly used into the MIMO backscatter system due to the following two reasons. 
For one thing, the channel model of the MIMO backscatter system is quite different from the conventional one and usually is modeled as the so-called \emph{dyadic backscatter channel} \cite{Boyer2014}, which makes it hard to formulate the considered security problem. 
For another thing, the passive backscatter system is typically resource-constrained and thus requires low-complexity algorithms in practice. 
Particularly, the trade-off between secrecy performance and computational complexity should be taken into account in the algorithm design of the MIMO backscatter system.
%Therefore, it is worth applying the MIMO and PLS techniques to backscatter communication systems to meet the requirement of a high and reliable data rate and strengthen the security in the backscatter wireless transmission.

Based on the above observations, in this paper we focus on solving the security issues of a MIMO RFID system from the perspective of PLS, wherein we take full account of the resource-constrained nature of RFID devices.
The novelty and main contributions of this paper can be summarized as follows: 
\begin{enumerate}[1)]
	\item The MIMO backscatter wireless communication is studied from the perspective of PLS for the first time and a noise-injection precoding strategy is proposed to strengthen the security of the system.
	\item %A comprehensive analysis of power allocation and the AN covariance matrix design is derived. Even though the formulated secrecy rate maximization problem is a non-convex matrix optimization problem, to a sequence of convex problems. 
	The \emph{alternating optimization} (AO) method and the \emph{sequential parametric convex approximation} (SPCA) method are, respectively, invoked to tackle the non-convex \emph{secrecy rate maximization} (SRM) problem. Furthermore, we show an interesting fact that the two methods are actually equivalent for our problem.
	\item Particularly, a custom-designed algorithm based on \emph{projected gradient} (PG) is proposed for fast implementation, which is especially beneficial to the resource-constrained RFID device.
	\item As a complement, the %more widely employed 
	case where the tag has only a single antenna is studied, and the global optimal solution can be obtained by one-dimensional search. % under some reasonable assumptions. 
	Moreover, a nullspace AN design in this case is further proposed, and it is shown in the simulations that this scheme obtains the secrecy rate which is close to the optimal one and enjoys an extremely low computational complexity. 
\end{enumerate}
%Numerical results show the significant advantages of the proposed noise-injection precoding scheme and fast algorithms, in strengthening the security of the MIMO backscatter wireless communication.

%5. Organization of this paper
The rest of this paper is organized as follows: In Section \ref{sec_model}, we present the system model and develop the formulation for the achievable secrecy rate of a MIMO RFID backscatter system. 
In Section \ref{sec_sol}, 
we focus on the SRM problem with a multi-antenna tag, and the equivalence of the AO and SPCA methods is analyzed. 
%the secrecy rate optimization problems in a multi-antenna tag case are formulated as matrix optimization problems.  
In Section \ref{sec_fast_algo}, we propose a fast algorithm to efficiently solve the SRM problem with a multi-antenna tag. 
The case where the tag equips with only a single antenna is studied in Section \ref{sec_single_ant}. 
Numerical simulations and analysis for the proposed schemes and algorithms are presented in Section \ref{sec_sim} before the conclusions drawn in Section \ref{sec_conclusion}.

\emph{Notations:} $\mathbf{A}^T$, $\mathbf{A}^H$, $\mathrm{det}(\mathbf{A})$ and $\mathrm{Tr}(\mathbf{A})$ represent the transpose, Hermitian transpose, determinant and trace of a matrix $\mathbf{A}$, respectively. $\mathbf{I}$ denotes an identity matrix. $\mathbf{A}\succeq\mathbf{0}$ means that $\mathbf{A}$ is a Hermitian positive semidefinite matrix. $\mathbb{E}\{\cdot\}$ and $(\cdot)^*$ denote the expectation and conjugate, respectively. $\mathbf{y}=[\mathbf{x}]^+$ means that $y_i=\max\{0,x_i\}$ for every $i$.
%$\mathrm{diag}\{\mathbf{x}\}$ and $\mathrm{diagm}\{\mathbf{A}\}$ denote a diagonal matrix whose diagonal elements come from a vector $\mathbf{x}$ and those of a matrix $\mathbf{A}$, respectively.
$\mathrm{diag}\{\mathbf{x}\}$ denotes a diagonal matrix with diagonal elements taken from vector $\mathbf{x}$.
$\mathbf{e}_i$ denotes a column vector whose the $i$-th element is 1 and 0 elsewhere. $\mathbf{x}\sim \mathcal{CN}(\mathbf{\bm{\mu}},\mathbf{\Sigma})$ means that $\mathbf{x}$ is a random vector following a complex circular Gaussian distribution with mean $\mathbf{\bm{\mu}}$ and covariance $\mathbf{\Sigma}$.
$\circ$ represents the Hadamard product.
%$\mathrm{Null}(\mathbf{H})$ denotes the nullspace of matrix $\mathbf{H}$.

\section{Secure MIMO Backscatter System Model}\label{sec_model}
\begin{figure}[t]
\centering
\includegraphics[width=3.0in]{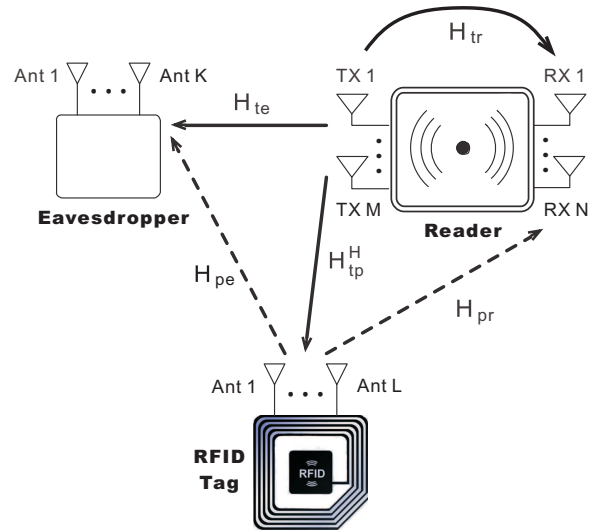}
\caption{%The MIMO backscatter wireless channel, where the signal at the receiver is cascaded of the $\mathbf{H}_{tp}$ and $\mathbf{H}_{pr}$ channels. Each tag antenna behaves as a pinhole, re-scattering the incident carrier wave from the reader. Eavesdropper receives the backscatter signal and the CW signal from channel $\mathbf{H}_{pe}$ and $\mathbf{H}_{te}$. The reader receives the self-interference signal from the channel $\mathbf{H}_{tr}$.
	The MIMO backscatter system, consisting of a RFID reader,
	a RFID tag, and a passive eavesdropper.
	}
\label{fig:1} \vspace{-3mm}
\end{figure}
Consider a RFID system consisting of a multi-antenna RFID reader with $M$ transmitting antennas and $N$ receiving antennas, a RFID tag with $L$ antennas, and a passive eavesdropper with $K$ receiving antennas as shown in Fig. \ref{fig:1}. For notational simplicity, we use the terms ``reader'' and ``tag'' as commonly used in RFID systems hereinafter.

The basic idea of realizing the RFID backscatter wireless communication is as follows: First, the reader transmits a continuous carrier wave (CW) to power up the tag. Then, the passive tag reflects back the CW signal by changing the different impedance loads of its antennas according to its stored secret information such as identification data. This procedure can be regarded as a backscatter modulation, where the secret information, to be transmitted from the tag to the reader, is modulated on the reflected CW signal by the tag. Finally, the reader extracts the desired information by estimating and decoding the echoed back signal after the backscatter modulation.
During the whole backscatter procedure, the reader continuously transmits a CW signal to power up the tag and concurrently receives the echoed back signal modulated by the tag, namely, it works in a full-duplex mode. 
Thus, the received self-interference or leaked signal directly from its transmitter to receiver needs to be canceled. 
Whereas in practice, it is difficult for the reader to perform this cancellation without any signal leakage. 
Thereby in this work, we use $\beta\in [0,1]$ as in \cite{Mukherjee2011} to denote the self-interference attenuation factor which reflects the capacity for the reader to cancel the interference from the transmitted signal in the received signal. 
Based on the procedure above, under the assumption that all the channels undergo slow frequency-flat fading, the received signal at the reader is given by\footnote{This signal model is first proposed in \cite{Ingram2001}, wherein self-interference is not considered and only spatial domain is involved. The model in \cite{Ingram2001} is further generalized to the space-time coding model in \cite{Boyer2014,Boyer2013,He2015} where temporal domain is also taken into account. In this paper, we focus on transmit optimization against eavesdropping only in spatial domain and thus adopt the signal model similar as that in \cite{Ingram2001}.}
\begin{align}
\mathbf{y}_r=\mathbf{H}_{pr}\mathbf{Q}\mathbf{H}_{tp}^H\mathbf{x}+\sqrt{\beta}\mathbf{H}_{tr}\mathbf{x'}+\mathbf{n}_r, \label{y_r}
\end{align}
where $\mathbf{x}\in \mathbb{C}^{M \times 1}$ is the signal transmitted by the reader\footnote{Since the reader receives its transmitted signals from both the backscatter and its self-interference channels at different time instants, we adopt $\mathbf{x'}$ in \eqref{y_r} to denote the signal received directly from the reader's self-interference channel and distinguish it from the signal received from the backscatter channel, i.e., $\mathbf{x}$. A similar notation will be also used in \eqref{y_e}.} with the total power constraint $\mathrm{Tr}(\mathbb{E}\{\mathbf{x}\mathbf{x}^H\})$ $ \leq P$.
$\mathbf{H}_{tp}^H\in \mathbb{C}^{L \times M}$, $\mathbf{H}_{pr}\in \mathbb{C}^{N \times L}$, and $\mathbf{H}_{tr}\in \mathbb{C}^{N \times M}$ represent the channel from the reader to the tag, the channel from the tag to the reader, and the self-interference channel at the reader, respectively.
$\mathbf{n}_r \sim \mathcal{CN}(\mathbf{0},\sigma_r^2\mathbf{I})$ is the additive white Gaussian noise (AWGN) at the reader.
$\mathbf{Q}\in\mathbb{C}^{L \times L}$ is the tag's information signaling matrix which represents the backscatter gain at the tag. 
The structure of $\mathbf{Q}$ describes the backscatter modulation by the tag, during which the RF tag absorbs and scatters radio signals by its $L$ antennas. This signaling matrix takes several different forms depending upon the physical implementation of the modulation circuitry and RF tag antennas\cite{Griffin2008}. 
In this paper, we consider a common scenario employed in \cite{Boyer2014,Ingram2001} that the signaling matrix $\mathbf{Q}$ takes the form of a diagonal matrix, given by
\begin{align}
\mathbf{Q}=\mathrm{diag}\{q_1, q_2, \cdots, q_L\}=\mathrm{diag}\{\mathbf{q}\}.
\end{align}
We assume that the elements in vector $\mathbf{q}$ are 
i.i.d. and $\mathbb{E}\{\mathbf{q}\mathbf{q}^H\}=\mathbf{I}$.

In this paper, we propose a noise-injection precoding scheme to create additional interference at the eavesdropper and thus to strengthen the security of the system. 
The transmit vector $\mathbf{x}$ takes the structure $\mathbf{x}=\mathbf{s}+ \mathbf{z}$, where $\mathbf{s} \in \mathbb{C}^{M\times1}$ is the CW signal to provide the tag with the power supply and we consider the conventional case as in uniform query\footnote{There exist two different query methods widely adopted in the existing literature: the conventional uniform query \cite{Boyer2014,Ingram2001,Boyer2013,He2014} and the newly developed unitary query adopted in the design of space-time coding for MIMO Backscatter RFID\cite{He2015,He2016}. In contrast to the diversity analysis in space-time coding where the message is coded over both space and time, here we focus on the transmit design against eavesdropping only in spatial domain and thus adopt conventional uniform query.} that the reader transmits a constant CW with the transmit power equally allocated among all its transmit antennas, i.e., $\mathbf{s} =\sqrt{P_s/M} \mathbf{1}_M$, where $P_s$ denotes the power allocated to the CW signal. $\mathbf{z} \in \mathbb{C}^{M\times1}$ is an AN vector, generated by the reader and injected into the transmitted CW signal, to interfere with the eavesdropper. Let $\mathbf{\Lambda}_z \triangleq \mathbb{E}\{\mathbf{z}\mathbf{z}^H\} \succeq \mathbf{0}$ denote the  spatial covariance matrix of the AN. The total power constraint at the reader now changes to
\begin{align}
P_s +\mathrm{Tr}(\mathbf{\Lambda}_z ) \leq P.  \label{p_constraint}
\end{align}

For the considered MIMO RFID channel, the reader uses the noisy observation of $\mathbf{H}_{pr}\mathbf{Q}\mathbf{H}_{tp}^H\mathbf{x}$ to estimate the unknown signal $\mathbf{q}$ with the CW signal $\mathbf{s}$. %Hence,  by modulating the different value of $\mathbf{Q}$, the tag's antenna array can spatially control the backscatter. 
We can rewrite $\mathbf{H}_{pr}\mathbf{Q}\mathbf{H}_{tp}^H\mathbf{x}$ as
\begin{equation}
\begin{split}
\mathbf{H}_{pr}\mathbf{Q}\mathbf{H}_{tp}^H\mathbf{x}&=\mathbf{H}_{pr}\diag\{\mathbf{H}_{tp}^H\mathbf{x}\}\mathbf{q}  \\
%=\mathbf{H}_{pr}\diag\{\mathbf{H}_{tp}^H\mathbf{s}+\mathbf{H}_{tp}^H\mathbf{z}\}\mathbf{q} \\
&=\sqrt{P_s}\mathbf{H}_{pr}\mathrm{diag}\{\sqrt{1/M}\mathbf{H}_{tp}^H\mathbf{1}_M\} \mathbf{q} \\
&~~~~~~+\mathbf{H}_{pr}\mathrm{diag}\{\mathbf{H}_{tp}^H\mathbf{z}\}\mathbf{q} .
\end{split}
\end{equation}
For notational simplification, let $\mathbf{D}_{tp}\triangleq\mathrm{diag}\{\sqrt{1/M}\mathbf{H}_{tp}^H\mathbf{1}_M\}$ and $\mathbf{F}_{tp}\triangleq\mathrm{diag}\{\mathbf{H}_{tp}^H\mathbf{z}\}$.
We can rewrite (\ref{y_r}) as
\begin{align}
\mathbf{y}_r=\sqrt{P_s}\mathbf{H}_{pr} \mathbf{D}_{tp} \mathbf{q}+\sqrt{\alpha}\mathbf{H}_{pr}\mathbf{F}_{tp}\mathbf{q}+\sqrt{\beta}\mathbf{H}_{tr}\mathbf{z'}+\mathbf{n}_r, \label{y_r_n}
\end{align}
where $\alpha \in [0,1]$, following from the setting in \cite{Saad2014}, is the attenuation factor which denotes how successful the reader is in canceling the backscattered AN. 
Note that the CW signal $\mathbf{s}'$, received directly from the self-interference channel $\mathbf{H}_{tr}$ at the reader, has been removed from \eqref{y_r_n} due to the fact that the standardized CW signal is commonly known and can be eliminated by the reader.
%Due to the difficulty to train and track channels and the resource-constrained nature for devices in RFID systems, we recognize that the noise attenuation for either the backscattered AN or the self-interference noise is actually difficult for the reader to perform. However, we do note that if such costs permits, the noise cancelation can be done via the standard signal processing technique which exploit the fact that the reader itself generated the noise, and thus, has prior knowledge of the noise signal $\mathbf{z}$. On the other hand, it is impossible for an external eavesdropper to perform the noise cancelation due to the absence of the prior knowledge about the random noise signal $\mathbf{z}$ generated by the reader.

%The main objective of the proposed noise injection scheme is to create additional interference at the eavesdropper during the reception of the backscatter signal from the tag. 
% the received signal model for
Similar to the reader, the received signal at the eavesdropper is given as
\begin{align}\label{y_e}
\mathbf{y}_e=\sqrt{P_s}\mathbf{H}_{pe} \mathbf{D}_{tp} \mathbf{q}+ \mathbf{H}_{pe}\mathbf{F}_{tp}\mathbf{q}+ \mathbf{H}_{te}\mathbf{z''}+\mathbf{n}_e,
\end{align}
where $\mathbf{H}_{pe} \in \mathbb{C}^{K \times L}$ and $\mathbf{H}_{te} \in \mathbb{C}^{K \times M}$ are the tag-eavesdropper and reader-eavesdropper channel matrices, respectively. 
$\mathbf{n}_e\sim \mathcal{CN}(\mathbf{0},\sigma_e^2\mathbf{I})$ is the AWGN at the eavesdropper.  $\mathbf{H}_{pe}\mathbf{F}_{tp}\mathbf{q}$ is the  backscattered AN signal modulated by the tag's information signal, while $\mathbf{H}_{te}\mathbf{z''}$ is the injected noise term received directly from the reader-eavesdropper channel. It should be noted that, unlike the reader, the eavesdropper cannot perform AN attenuation due to the absence of the prior knowledge about the random AN signal $\mathbf{z}$ transmitted by the reader. 
However, the CW signal received directly from the reader-eavesdropper channel can be eliminated by the eavesdropper due to the fact that the standardized CW signal is commonly known\cite{Saad2014}. That is why this signal term does not appear in \eqref{y_e}.

The achievable secrecy rate for a given secure transmission scheme determines the performance limit of PLS.
Unfortunately, the exact expression of the secrecy rate here is difficult to obtain due to the non-Gaussian distribution of the combined signal and AN terms $\mathbf{H}_{pr}\mathbf{F}_{tp}\mathbf{q}$ and $\mathbf{H}_{pe}\mathbf{F}_{tp}\mathbf{q}$ received at the reader and the eavesdropper, respectively. However, following the similar method adopted in \cite{Saad2014}, we regard these terms as interference and obtain an approximation of the achievable secrecy rate given by \cite{Oggier2011}
\begin{subequations} \label{Cs_C}
	\begin{align}
	&C_s = [C_r-C_e]^+, \label{Cs} \\
	&C_r\approx \log_2\det(\mathbf{I}_N+ P_s  \mathbf{H}_{pr} \mathbf{D}_{tp}\mathbf{D}_{tp}^H\mathbf{H}_{pr}^H \mathbf{R}_r^{-1}), \label{a_rates}\\
	&C_e\approx \log_2\det(\mathbf{I}_K + P_s  \mathbf{H}_{pe} \mathbf{D}_{tp}\mathbf{D}_{tp}^H\mathbf{H}_{pe}^H \mathbf{R}_e^{-1}), \label{a_rates1}
	\end{align}
\end{subequations}
where the covariance matrices of the interference and noise are given by
\begin{subequations}  \label{IN_cov}
\begin{align}
&\mathbf{R}_r=\alpha\mathbf{H}_{pr}\mathbb{E}_{\mathbf{z}}[\mathbf{F}_{tp}\mathbf{F}_{tp}^H]\mathbf{H}_{pr}^H
+\beta\mathbf{H}_{tr}\mathbf{\Lambda}_z\mathbf{H}_{tr}^H
+\sigma_r^2\mathbf{I}_N,  \label{Rr}\\
&\mathbf{R}_e=\mathbf{H}_{pe}\mathbb{E}_{\mathbf{z}}[\mathbf{F}_{tp}\mathbf{F}_{tp}^H]\mathbf{H}_{pe}^H
+\mathbf{H}_{te}\mathbf{\Lambda}_z\mathbf{H}_{te}^H
+\sigma_e^2\mathbf{I}_K,
\end{align}
\end{subequations}
with
\begin{align*}
\mathbb{E}_{\mathbf{z}}[\mathbf{F}_{tp}\mathbf{F}_{tp}^H]&=
\mathrm{diag}\big\{\mathbb{E}[\mathbf{H}_{tp}^H\mathbf{z} \circ (\mathbf{z}^H\mathbf{H}_{tp})^T]\big\} \\
&=\mathbb{E}[\mathbf{H}_{tp}^H\mathbf{z}  \mathbf{z}^H\mathbf{H}_{tp}] \circ \mathbf{I} \\
&=(\mathbf{H}_{tp}^H\mathbf{\Lambda}_z\mathbf{H}_{tp}) \circ \mathbf{I} \\
&=\sum_{i=1}^L\left(\mathbf{e}_i^T\mathbf{H}_{tp}^H\mathbf{\Lambda}_z\mathbf{H}_{tp}\mathbf{e}_i\right)\left(\mathbf{e}_i\mathbf{e}_i^T\right).
\end{align*}
%Thus, the power allocation between the AN and the CW as well as the covariance matrix of AN $\mathbf{\Lambda}_z$ will be designed to maximize the achievable secrecy rate with a total transmit power constraint.

%Note that when the antenna number at the transmitter of the reader is not large enough, there does not exist a nullspace to transmit AN such that the reader itself will not receive it. 
%Thus, in this situation 
From \eqref{Cs_C} and \eqref{IN_cov}, to maximize the achievable secrecy rate in \eqref{Cs} under the total power constraint in \eqref{p_constraint}, the power allocation between the AN and the CW as well as the covariance matrix of AN $\mathbf{\Lambda}_z$ needs to be carefully designed since the jamming signal from the reader degrades the performance of both the reader itself and the eavesdropper. 
The case where the tag has multiple antennas will be studied in the next two sections, while the %more widely employed 
scenario where a single antenna is employed at the tag will be investigated in Section \ref{sec_single_ant}.

\section{Multi-Antenna Tag} \label{sec_sol}
In this section, we consider the case where the tag equips with multiple antennas. The considered problems are first formulated as matrix optimization problems in Section \ref{sec_formulation}, and then the AO and SPCA methods are introduced to solve the SRM problem in Section \ref{sec_ao} and Section \ref{sec_spca}, respectively. The equivalence of the two methods in the considered problem will be illustrated in Section \ref{sec_equivalent}.

\subsection{Problem Formulation} \label{sec_formulation}
%In this subsection, we formulate our PLS problems in a MIMO backscatter wireless system where the tag has multiple antennas as two optimization problems. 
To facilitate analysis, we first recast the secrecy rate in \eqref{Cs} as
\begin{align}
C_s(P_s,\mathbf{\Lambda}_z) &= \ln\det(\mathbf{R}_r+P_s \mathbf{A} )+\ln\det(\mathbf{R}_e) \notag\\
&~~~~ -\ln\det(\mathbf{R}_r) -\ln\det(\mathbf{R}_e +P_s \mathbf{B}),  \label{Cs1}
\end{align}
where $\mathbf{A}\triangleq \mathbf{H}_{pr} \mathbf{D}_{tp}\mathbf{D}_{tp}^H\mathbf{H}_{pr}^H$ and $\mathbf{B}\triangleq\mathbf{H}_{pe} \mathbf{D}_{tp}\mathbf{D}_{tp}^H\mathbf{H}_{pe}^H$.
Our aim is to maximize the achievable secrecy rate in \eqref{Cs1} under the total transmit power constraint in \eqref{p_constraint}. 
Mathematically, this %\emph{secrecy rate maximization} (
SRM problem can be formulated as follows:%a matrix optimization problem as follows:
%\begin{subequations}
	\begin{align}\label{general_OP}
	\max_{P_s,\mathbf{\Lambda}_z}~~&C_s(P_s,\mathbf{\Lambda}_z)  ~~~~~
	\st ~~(P_s,\mathbf{\Lambda}_z) \in \mathcal{C},
	\end{align}
%\end{subequations}
where the feasible set is defined as
\begin{align}\label{constraint}
\mathcal{C} \triangleq \{(P_s,\mathbf{\Lambda}_z)~|~ P_s+\Tr(\mathbf{\Lambda}_z ) \leq P,P_s \geq 0, \mathbf{\Lambda}_z\succeq\mathbf{0}\}.
\end{align}

As discussed in \cite{Wang2015,Li2013}, a simplified and special design of the general AN scheme in \eqref{general_OP} is the so-called nullspace AN scheme where the transmitted AN lies in the nullspace of the legitimate user's channel. 
In our MIMO backscatter model, from \eqref{Rr} we know that the reader receives two altered copies of its transmitted AN from the backscatter channel and the self-interference channel, respectively. When both the AN copies cannot be perfectly eliminated at the same time and the number of transmit antennas at the reader is adequate to perform the nullspace AN precoding, namely $M>L$ or $M>N$, the AN transmitted by the reader can be designed to lie in the nullspace of the reader-tag channel $\mathbf{H}_{tp}^H$ or the self-interference channel $ \mathbf{H}_{tr} $, respectively.
To be specific, we can construct the injected AN %at the transmitter of the reader 
as $\mathbf{z}=\mathbf{V}\mathbf{w}$, and now we have
\begin{align}
\mathbf{\Lambda}_z(\mathbf{W})=\mathbf{V}\mathbf{W}\mathbf{V}^H, \label{nullcs}
\end{align}
where $\mathbf{V}$ contains all the right singular vectors of $\mathbf{H}_{tp}^H$ or $ \mathbf{H}_{tr} $ corresponding to zero singular values, and $\mathbf{W}=\mathrm{E}\{\mathbf{w}\mathbf{w}^H\}$ is an $(M-L)\times(M-L)$ or $(M-N)\times(M-N)$ positive semidefinite matrix to be optimized, respectively. 
%Under this nullspace AN constraint, we have $\mathbf{F}_{tp}=\mathbf{0}$ or $ \mathbf{H}_{tr}\mathbf{z}=0 $, respectively, and thus the receiving (SINR) at the reader can be improved. 
%Apparently, this scheme is suboptimal due to the restricted structure of the AN covariance matrix, whereas this scheme has a lower computational complexity compared with problem \eqref{general_OP} since the degree of freedom at the transmitter of the reader is reduced.
%both SNRs at the reader and eavesdropper are improved. But intuitively, it can be beneficial to perform this nullspace AN precoding scheme when the attenuation factor $\alpha$ is large. This observation will be verified by numerical simulation in Section \ref{sec_sim}. 
%In addition, since the dimension of the variable to be optimized has been reduced, 
Mathematically, the nullspace SRM problem is formulated as follows:
%\begin{subequations}
	\begin{align}\label{null_BF}
	\max_{P_s,\mathbf{W}}~~&C_s(P_s,\mathbf{\Lambda}_z(\mathbf{W}))  ~~~~~
	\st ~~(P_s,\mathbf{W}) \in \mathcal{C}.  %\label{null_BF_cs}
	\end{align}
%In the simulation, we will use this nullspace AN method as a benchmark to evaluate the performance and computational complexity of our proposed  algorithms.  
Since \eqref{null_BF} is just a degenerate form of the SRM problem in \eqref{general_OP}, we will focus on solving \eqref{general_OP} hereinafter.

The SRM problem in \eqref{general_OP} is non-convex and difficult to tackle due to the non-concave property of the term $-\ln\det(\cdot)$ in the objective function in \eqref{Cs1}. 
The AO and SPCA methods are the two methods widely exploited in tackling non-convex matrix optimization problems \cite{Li2013a,Marks1978,Beck2010}, and it will be interesting to see in the sequel that the two methods are actually equivalent under our SRM problem. Therefore, our approach is to reformulate the non-concave term to a concave one by exploiting the AO and SPCA methods. The two methods will be, respectively, introduced in the next two subsections.
%To handle this, our approach is to reformulate the non-concave term to a concave one by exploiting the AO method or the SPCA method. The two methods will be, respectively, introduced in the next two subsections.

\subsection{AO Method for SRM} \label{sec_ao}
The main idea of the AO method is to exploit the coordinate-wise convexity property of the non-convex problem where optimization over two subsets of variables is non-convex, but optimization with respect to (w.r.t.) one while fixing the other is convex. 
%Thus, in this case the AO method can be used to iteratively solve the two convex optimization problems to obtain a solution of the original non-convex problem. 
To re-express the general SRM problem in \eqref{general_OP} as a form that can be tackled by the AO method, we first introduce the following lemma.
\begin{lemma}\cite{Jose2011}  \label{lem:1}
Let $\mathbf{X}\in\mathbb{C}^{N \times N}$ and $\mathbf{X}\succ\mathbf{0}$, then the function $ -\ln\det(\mathbf{X}) $ can be equivalently rewritten by importing an auxiliary variable $ \mathbf{S} \in\mathbb{C}^{N \times N}$ as
\begin{align}
 -\ln\det(\mathbf{X}) =\max_{\mathbf{S}\succeq\mathbf{0}}-\mathrm{Tr}(\mathbf{S}\mathbf{X})+\ln\det (\mathbf{S})+N, \label{lem1}
\end{align}
and the right-hand side of \eqref{lem1} has the optimal solution $\mathbf{S}^\star=\mathbf{X}^{-1}$ in a closed form.
\end{lemma}

From Lemma \ref{lem:1}, one can easily see that the non-concave term can be changed to a linear (and thus concave) term w.r.t. the original optimization variable by adding an auxiliary variable. By applying Lemma \ref{lem:1} to the objective function in \eqref{Cs1} via setting $\mathbf{X}_0=\mathbf{R}_r$ and $\mathbf{X}_1=\mathbf{R}_e + P_s \mathbf{B}$, we have the following equivalent formulation of problem \eqref{general_OP}:
\begin{equation}
\begin{split}
\max_{\substack{P_s,\mathbf{\Lambda}_z,\mathbf{S}_0,\mathbf{S}_1}}~~&\ln\det(\mathbf{R}_r+P_s \mathbf{A} ) +\ln\det(\mathbf{R}_e) \\
&~~~~-\mathrm{Tr}(\mathbf{S}_0\mathbf{R}_r)+\ln\det(\mathbf{S}_0)  \\
&~~~~-\mathrm{Tr}[\mathbf{S}_1(\mathbf{R}_e +P_s \mathbf{B})]+\ln\det(\mathbf{S}_1)  \\
\st ~~&(P_s,\mathbf{\Lambda}_z) \in \mathcal{C},~\mathbf{S}_0\succeq\mathbf{0},~\mathbf{S}_1\succeq\mathbf{0}.    \label{general_OP1}
\end{split}
\end{equation}
Note that we have dropped the constant $K+N$ in the objective of problem  \eqref{general_OP1} for simplicity. The equivalent problem in \eqref{general_OP1} is non-convex w.r.t. $(P_s,\mathbf{\Lambda}_z,\mathbf{S}_0,\mathbf{S}_1)$. However, it is not hard to see that problem \eqref{general_OP1} is convex w.r.t. either $(P_s,\mathbf{\Lambda}_z)$ or $(\mathbf{S}_0,\mathbf{S}_1)$ while fixing the other. %, fixing the other optimization variable. %\cite{Li2013a}. 
We can exploit this coordinate-wise convexity property to use the AO method to solve the problem. To be specific, let $(P_s^n,\mathbf{\Lambda}_z^n,\mathbf{S}_0^n,\mathbf{S}_1^n)$ denote the solution obtained at the $n$-th AO iteration. We iteratively use the values at the $(n-1)$-th iteration to obtain the ones at the $n$-th iteration by alternatingly solving the following two optimization problems for $n=1,2,\cdots$,
\begin{subequations}
	\begin{align}
	(\mathbf{S}_0^n,\mathbf{S}_1^n)&=\arg\max_{\mathbf{S}_0,\mathbf{S}_1\succeq\mathbf{0}}~\ln\det(\mathbf{S}_0) -\mathrm{Tr}(\mathbf{S}_0\mathbf{R}_r^{n-1}) \nonumber \\
	&~~+\ln\det(\mathbf{S}_1)-\mathrm{Tr}[\mathbf{S}_1(\mathbf{R}_e^{n-1} +P_s^{n-1} \mathbf{B})],  \label{S01}\\
	(P_s^n,\mathbf{\Lambda}_z^n)&=\arg\max_{(P_s,\mathbf{\Lambda}_z) \in \mathcal{C}}~\ln\det(\mathbf{R}_r+P_s \mathbf{A} )+\ln\det(\mathbf{R}_e) \nonumber \\
	&~~-\mathrm{Tr}(\mathbf{S}_0^n\mathbf{R}_r)-\mathrm{Tr}[\mathbf{S}_1^n(\mathbf{R}_e +P_s \mathbf{B})].% \nonumber \\
	%&~~~~~~~~~~\st ~~(P_s,\mathbf{\Lambda}_z) \in \mathcal{C}. 
	\label{Lambda}
	\end{align}
\end{subequations}

From Lemma \ref{lem:1}, the optimal solution to problem \eqref{S01} takes a closed form and can be obtained by
%\begin{subequations}
\begin{align}\label{S01_solu}
&\mathbf{S}_0^n=(\mathbf{R}_r^{n-1})^{-1}, ~~~
\mathbf{S}_1^n=(\mathbf{R}_e^{n-1} +P_s^{n-1} \mathbf{B})^{-1}.
\end{align}
%\end{subequations}
%However, problem \eqref{Lambda} is quite complex and hard to get a closed form solution. Here we use \verb"CVX"\cite{7}, a general-purpose convex optimization solver, to obtain a numerical solution.
Problem \eqref{Lambda} is convex, and thus can be numerically solved. 
The solution to our SRM problem can be obtained by iteratively calculating \eqref{S01_solu} and solving problem \eqref{Lambda} until the corresponding secrecy rate fulfills the given accuracy requirement. 
%We summarize the procedure to solve the SRM problem by the AO method in Algorithm \ref{alg:AO}.
%\begin{algorithm}
%\caption{AO Method for SRM}\label{alg:AO}
%\begin{algorithmic}[1]
%\REQUIRE $P$, $n=1$, $(P_s^0,\mathbf{\Lambda}_z^0) \in \mathcal{C}$, $\epsilon>0$;
%%\ENSURE $P_s^\star,~\mathbf{\Lambda}_z^\star$;
%\WHILE{$C_s(P_s^n,\mathbf{\Lambda}_z^n)$ not converge for the tolerance $\epsilon$}
%\STATE Update $(\mathbf{S}_0^n,\mathbf{S}_1^n)$ by \eqref{S01_solu};
%\STATE Update $(P_s^n,\mathbf{\Lambda}_z^n)$ by solving \eqref{Lambda};
%\STATE $n=n+1$;
%\ENDWHILE
%\RETURN $P_s^\star=P_s^n,~\mathbf{\Lambda}_z^\star=\mathbf{\Lambda}_z^n$;
%\end{algorithmic}
%\end{algorithm}

As a basic result of AO, the method proposed above produces non-descending objective values. More specifically, we have $C_s(P_s^0,\mathbf{\Lambda}_z^0)\leq C_s(P_s^1,\mathbf{\Lambda}_z^1)\leq \cdots \leq C_s(P_s^n,\mathbf{\Lambda}_z^n)$\cite{Li2013a}. In addition, we will show by Theorem \ref{th_converge} in Section \ref{sec_equivalent} that the sequence $\{P_s^n,\mathbf{\Lambda}_z^n\}$ generated by the AO method converges to a Karush-Kuhn-Tucker (KKT) point of the original problem in \eqref{general_OP}.

\subsection{SPCA Method for SRM} \label{sec_spca}
As an alternative way, we can also use the SPCA method to solve the SRM problem in \eqref{general_OP}. The basic idea of the SPCA method is to  approximate a non-convex problem by a sequence of convex problems. In each convex problem, every non-convex constraint is replaced by an appropriate inner but convex one. %Under the appropriate conditions on the inner convex approximation, a monotone convergence to a KKT point is established and the rigorous proof can be found in \cite[Proposition 3.2]{Beck2010}. 
Generally, the convergence rate of the SPCA method is fast. More details about SPCA can be found in \cite{Marks1978,Beck2010}.

To apply the SPCA method, we first transform the non-concave terms in the objective function in \eqref{Cs1} to the inner-approximated but concave ones. Note that the non-concave terms in the objective function are actually convex w.r.t. the optimization variable. Here the first-order Taylor's series approximation can be used as a global underestimator of the convex function $-\ln\det(\mathbf{X})$ \cite[p. 69]{Boyd2004}. More specifically, the approximation for the function $-\ln\det(\mathbf{X})$ w.r.t. $ \mathbf{X} \succ 0 $ around $ \mathbf{X}_0 $ is given by
\begin{align}\label{log_es}
    -\ln\det(\mathbf{X}) \geq -\ln\det(\mathbf{X}_0)-\mathrm{Tr}[\mathbf{X}_0^{-1}(\mathbf{X}-\mathbf{X}_0)].
\end{align}
From the right-hand side of \eqref{log_es}, one can see that non-concave terms can be changed to linear (and thus concave) ones by exploiting this method.

By applying the approximation in \eqref{log_es} centering around the point $(P_s^{n-1},\mathbf{\Lambda}_z^{n-1})$ to non-concave terms in the objective function in \eqref{Cs1}, we obtain \eqref{f_approxi} at the top of the next page\addtocounter{equation}{1}
%shown at the top of the next page, %!!!!!!!!!!!!!!!!!!!!
where the superscript $n-1$ refers to the optimal solution obtained at the $(n-1)$-th iteration. 
In each iteration we perform the approximation centering around the optimal solution obtained at the previous iteration. At the $n$-th iteration, the original non-convex SRM problem in \eqref{general_OP} can be locally approximated by the following convex optimization problem for $n=1,2,\cdots$,
\begin{align}\label{general_OP2}
&(P_s^n,\mathbf{\Lambda}_z^n)=\arg\max_{(P_s,\mathbf{\Lambda}_z) \in \mathcal{C}}~f_0(P_s,\mathbf{\Lambda}_z) \notag\\
&-f_1(P_s,\mathbf{\Lambda}_z,P_s^{n-1},\mathbf{\Lambda}_z^{n-1})-f_2(P_s,\mathbf{\Lambda}_z,P_s^{n-1},\mathbf{\Lambda}_z^{n-1}),
% \nonumber \\
%\st ~~&(P_s,\mathbf{\Lambda}_z) \in \mathcal{C},
\end{align}
where
\begin{align}\label{f0}
f_0(P_s,\mathbf{\Lambda}_z)\triangleq\ln\det(\mathbf{R}_r+P_s \mathbf{A} )+\ln\det(\mathbf{R}_e)
\end{align}
with $f_1(P_s,\mathbf{\Lambda}_z,P_s^{n-1},\mathbf{\Lambda}_z^{n-1})$ and $f_2(P_s,\mathbf{\Lambda}_z,P_s^{n-1},\mathbf{\Lambda}_z^{n-1})$ defined in \eqref{f_approxi}.
%Again, we can use \verb"CVX" to get the numerical solution to the problem in \eqref{general_OP2}.
The solution to our SRM problem can be obtained by iteratively solving problem \eqref{general_OP2} until the corresponding secrecy rate fulfills the given accuracy requirement.

\setcounter{mytempeqncnt}{\value{equation}}
\setcounter{equation}{18}%!!!!!!!!!!!!!!!!!!!!!!!!!!!!!!!!!!!!!!!!!!!!!!!
\begin{figure*}[!t]
	%\vspace*{4pt}
	\begin{subequations}\label{f_approxi}
		\begin{align}
			&\ln\det(\mathbf{R}_r) \leq f_1(P_s,\mathbf{\Lambda}_z,P_s^{n-1},\mathbf{\Lambda}_z^{n-1})\triangleq \ln\det(\mathbf{R}_r^{n-1}) \nonumber\\
			&~~~
			+\mathrm{Tr}\left\{(\mathbf{R}_r^{n-1})^{-1}\left[\alpha\mathbf{H}_{pr}\left(\left(\mathbf{H}_{tp}^H(\mathbf{\Lambda}_z-\mathbf{\Lambda}_z^{n-1})\mathbf{H}_{tp}\right) \circ \mathbf{I}\right)\mathbf{H}_{pr}^H
			+\beta\mathbf{H}_{tr}(\mathbf{\Lambda}_z-\mathbf{\Lambda}_z^{n-1})\mathbf{H}_{tr}^H\right] \right\}, \label{f_approxi_Rr}\\
			&\ln\det(\mathbf{R}_e +P_s \mathbf{B}) \leq f_2(P_s,\mathbf{\Lambda}_z,P_s^{n-1},\mathbf{\Lambda}_z^{n-1}) \triangleq \ln\det(\mathbf{R}_e^{n-1} +P_s^{n-1} \mathbf{B})
			+\mathrm{Tr}\big\{(\mathbf{R}_e^{n-1} +P_s^{n-1} \mathbf{B})^{-1}
			\nonumber\\
			&~~~\left[\mathbf{H}_{pe}\left(\left(\mathbf{H}_{tp}^H(\mathbf{\Lambda}_z-\mathbf{\Lambda}_z^{n-1})\mathbf{H}_{tp}\right) \circ \mathbf{I}\right)\mathbf{H}_{pe}^H
			+\mathbf{H}_{te}(\mathbf{\Lambda}_z-\mathbf{\Lambda}_z^{n-1})\mathbf{H}_{te}^H+(P_s-P_s^{n-1} )\mathbf{B} \right]\big\}.
		\end{align}
	\end{subequations}
	\hrulefill
\end{figure*}
\setcounter{equation}{\value{mytempeqncnt}}

%The procedure to solve the SRM problem by the SPCA method is shown in Algorithm \ref{alg:SPCA}.
%\begin{algorithm}
%\caption{SPCA Method for SRM}\label{alg:SPCA}
%\begin{algorithmic}[1]
%\REQUIRE $P$, $n=1$, $(P_s^0,\mathbf{\Lambda}_z^0) \in \mathcal{C}$, $\epsilon>0$;
%%\ENSURE $P_s^\star,~\mathbf{\Lambda}_z^\star$;
%\WHILE{$C_s(P_s^n,\mathbf{\Lambda}_z^n)$ not converge for the tolerance $\epsilon$}
%\STATE Update $(P_s^n,\mathbf{\Lambda}_z^n)$ by solving \eqref{general_OP2};
%\STATE $n=n+1$;
%\ENDWHILE
%\RETURN $P_s^\star=P_s^n,~\mathbf{\Lambda}_z^\star=\mathbf{\Lambda}_z^n$;
%\end{algorithmic}
%\end{algorithm}

%As a basic result of SPCA, after each iteration the objective function is non-decreasing, and the convergence to a local optimum is guaranteed because the achievable secrecy rate $C_s$ is up-bounded for the given transmit power $P$.
Like the AO method in the last subsection,  the sequence of objective values produced by the SPCA method is non-descending. Moreover,
we will show by Theorem \ref{th_converge} in the next subsection that the sequence $\{P_s^n,\mathbf{\Lambda}_z^n\}$ generated by the SPCA method converges to a KKT point of the original SRM problem in  \eqref{general_OP}.

\begin{remark}
As seen in Section \ref{sec_formulation}, the nullspace SRM problem in \eqref{null_BF} is just a degenerate form of the SRM problem in  \eqref{general_OP} when the nullspace AN constraint in \eqref{nullcs} is imposed. Thus, once the general SRM problem in \eqref{general_OP} is solved by the AO or SPCA method, the  nullspace SRM problem in \eqref{null_BF} can be easily solved in a similar way.% by imposing the nullspace AN constraint in \eqref{nullcs} with the optimization variable changed to $\mathbf{W}$.
\end{remark}
\begin{remark}
Generally, the AO and SPCA methods are two totally different methods and either the AO or SPCA method \emph{individually} will give a solution. However, it is interesting to see that the two methods are actually \emph{equivalent} for our SRM problem in the sense that the two methods can equivalently lead to the same optimization problem. 
We will analyze this equivalence in the next subsection.
\end{remark}

\subsection{Equivalence, Convergence, and Complexity Analyses}\label{sec_equivalent}
%We now illustrate that the AO and SPCA methods are actually equivalent for our SRM problem due to the fact that the two methods can be equivalently transformed to the same optimization problem.

As for the AO method, the equivalence can be verified by putting \eqref{S01_solu} in \eqref{Lambda}, and then we obtain the following convex optimization problem at the $n$-th iteration:
\begin{align}\label{general_OP3}
(P_s^n,\mathbf{\Lambda}_z^n)=&\arg\max_{(P_s,\mathbf{\Lambda}_z) \in \mathcal{C}}~g(P_s,\mathbf{\Lambda}_z,P_s^{n-1},\mathbf{\Lambda}_z^{n-1}), 
%\nonumber \\
%~~~~~~\st ~~(P_s,\mathbf{\Lambda}_z) \in \mathcal{C},
\end{align}
where
\begin{align}\label{objective}
&g(P_s,\mathbf{\Lambda}_z,P_s^{n-1},\mathbf{\Lambda}_z^{n-1}) \triangleq f_0(P_s,\mathbf{\Lambda}_z)-\mathrm{Tr}((\mathbf{R}_r^{n-1})^{-1}\mathbf{R}_r) \notag\\
&~~~~~~~~~~~-\mathrm{Tr}[(\mathbf{R}_e^{n-1} +P_s^{n-1} \mathbf{B})^{-1}(\mathbf{R}_e +P_s \mathbf{B})].
\end{align}
It is not hard to see that the objective function of problem \eqref{general_OP2} obtained by the SPCA method and \eqref{objective} only differ in some added terms that are irrespective of our optimization variable $(P_s,\mathbf{\Lambda}_z)$, and thereby have no effect on the optimization result. This leads to the equivalence of the two methods for the considered SRM problem.

The main reasons for this equivalence are summarized as follows:
\begin{enumerate}[1)]
	\item As for the SPCA method, the terms $\mathbf{R}_r$ and $\mathbf{R}_e +P_s \mathbf{B}$ in $-\ln\det(\cdot)$ of the objective function in \eqref{Cs1} are linear w.r.t. the optimization variable $(P_s,\mathbf{\Lambda}_z)$. This linear property makes the terms $-\ln\det(\cdot)$ in the objective function convex, and thus the approximation of  the first-order Taylor's series in \eqref{log_es} can be applicable. 
%	Under the condition that the inner approximate function in the SPCA method is chosen as the first-order Taylor's series, this linear property combining with the differential property of $\ln\det(\cdot)$ 
	The choice of this approximation in the SPCA method makes the equivalence between problem \eqref{general_OP2} obtained by the SPCA method and problem \eqref{general_OP3} possible.
%This can be shown more specifically by the derivation of \eqref{f_approxi_Rr}. Because the differential of $\ln\det(\mathbf{R}_r)$ is $\mathrm{Tr}\{\mathbf{R}_r^{-1}d\mathbf{R}_r\}$ and $\mathbf{R}_r$ is a linear combine of $\mathbf{\Lambda}_z$, one can easily get the term $\mathrm{Tr}((\mathbf{R}_r^{n-1})^{-1}\mathbf{R}_r)$ in \eqref{general_OP3} by applying \eqref{log_es} and eliminating some added constant terms such like the terms including $\mathbf{\Lambda}_z^{n-1}$.
	\item Concerning the AO method, Lemma \ref{lem:1} plays an important role in this equivalence. It should be noted that the specific way to apply the AO method actually depends largely on the form of the considered problem. In our case, it is the non-concave term $-\ln\det(\cdot)$ that makes Lemma \ref{lem:1} applicable, and then the AO method can be employed to lead us to problem \eqref{general_OP3}.
	\item It can be observed that \eqref{lem1} in Lemma \ref{lem:1} of the AO method is actually closely related to \eqref{log_es} in the SPCA method. This is because \eqref{lem1} means that
	\begin{align}
	&-\ln\det(\mathbf{X}) \geq -\mathrm{Tr}(\mathbf{S}\mathbf{X})+\ln\det (\mathbf{S})+N, \notag\\
	&~~~~~~~~~~~~~~~~~~~~~~~~\mathrm{for~any}~\mathbf{X}\succ 0,~\mathbf{S}\succeq 0 \notag \\%,~=~\mathrm{stisfies when}~\mathbf{S}=\mathbf{X}^{-1} \\
	\xLongrightarrow{\mathbf{S}=\mathbf{X}_0^{-1}} &
	-\ln\det(\mathbf{X}) \geq -\mathrm{Tr}[\mathbf{X}_0^{-1}(\mathbf{X}-\mathbf{X}_0)]-\ln\det (\mathbf{X}_0), \notag\\
	&~~~~~~~~~~~~~~~~~~~~~~~~\mathrm{for~any}~\mathbf{X}\succ 0,
	\end{align}
	which is exactly \eqref{log_es}.
\end{enumerate}

Since we have shown that both the AO and SPCA methods for our SRM problem are equivalent to iteratively solving the convex optimization problem in \eqref{general_OP3}, the two methods actually have the same convergence result and computational complexity. 
We first give the convergence result in the following theorem.
\begin{theorem}\label{th_converge}
Both the AO and SPCA methods produce non-descending achievable secrecy rates and converge to a KKT point of the original SRM problem in \eqref{general_OP}.
\end{theorem}
\begin{IEEEproof}
The proof is given in Appendix \ref{app_converge}.
\end{IEEEproof}

At each AO or SPCA iteration, the convex optimization problem in \eqref{general_OP3} can be solved by using a general-purpose convex optimization toolbox, such as \verb"CVX"\cite{Grant2011}, to obtain a numerical solution. 
This computational complexity can be approximated by the complexity of solving a standard semidefinite program (SDP) problem through the interior point method, though problem \eqref{general_OP3} is not a standard SDP problem.
Problem \eqref{general_OP3} has $M^2$ independent real and imaginary parts in the Hermitian matrix $ \L $. Let $J$ denote the total number of the AO or SPCA iteration. % whose typical value is ten when the solution accuracy at the algorithm's termination is $10^{-3}$. 
Then the complexity cost of the AO or SPCA method is at most $O(JM^7)$\cite{Ye1997}, while the cost reduces to at most $O(J(M-L)^7)$ or $O(J(M-N)^7)$ for the two nullspace SRM problems in \eqref{null_BF}. These computational complexity costs are usually much less when the modern SDP solver like SeDuMi\cite{Sturm1999} in \verb"CVX" is employed. 
%Actual runtime complexity usually scales far slower with the antenna numbers than these worst-case bounds, and the solving of the problems usually takes several seconds on a typical personal computer.

%\subsection{Bisection Based Method for Power Minimization} \label{sec_bi}
%At this section, the power minimization (PM) problem in \eqref{power_min} will be solved based on bisection method and the method proposed in solving the SRM problem in \eqref{general_OP}.
%
%Intuitively, the optimal $C_s(P)$ in the SRM problem is an increasing function about the total power $P$, i.e. $C_s(P)$ decreases with a decrease in $P$. It can be easily shown by contradiction and observing that the inequality in \eqref{power_min_geq} always holds with equality to achieve the minimum power. Hence, from this observation one can easily see that the optimal $P(r)$ in the PM problem is an inverse function of the optimal $C_s(P)$ in the SRM problem. This monotonic property inspires us to perform a bisection search to $P$ in the SRM problem to solve the PM problem, and the procedure is shown in Algorithm \ref{alg:1}.
%\begin{algorithm}
%\caption{Bisection Based Method for PM}\label{alg:1}
%\begin{algorithmic}[1]
%\REQUIRE $P_u>P_l>0,~r,~\epsilon>0$;
%\ENSURE $P^\star$ satisfying $C_s(P^\star)\simeq r$;
%\WHILE{$|P_u-P_l|>\epsilon$}
%\STATE $P_{mid}=(P_u-P_l)/2$;
%\STATE Use the AO or SPCA method to solve $C_s(P_{mid})$;
%\IF{$C_s(P_{mid})>r$}
%\STATE $P_u=P_{mid}$;
%\ELSE
%\STATE $P_l=P_{mid}$;
%\ENDIF
%\ENDWHILE
%\RETURN $P^\star=(P_u+P_l)/2$;
%\end{algorithmic}
%\end{algorithm}

\section{A Fast PG Algorithm for SRM with a Multi-Antenna Tag}\label{sec_fast_algo}
As we have indicated, the RFID system is typically resource-constrained compared with the conventional wireless communication system. Thus, it particularly requires low-complexity algorithms in practice. 
%From Section \ref{sec_equivalent}, we know that the AO and SPCA methods are actually equivalent and the main idea to solve the SRM problem with a multi-antenna tag is to solve a series of inner convex optimization problems in \eqref{general_OP3}. As we have mentioned, the convex optimization package \verb"CVX" can be exploited to solve the inner problem in  \eqref{general_OP3}. However, the procedure to use \verb"CVX" to solve problem \eqref{general_OP3} is time-costing and low efficient due to the general-purpose property of the solver. Hence, 
In this section, by exploiting the characteristic of the feasible set of the SRM problem in \eqref{general_OP}, we develop a fast and low-complexity algorithm based on projected gradient (PG) to replace the inefficient \verb"CVX" while solving the convex problem in \eqref{general_OP3}.

Let $(P_s^{n,k},\mathbf{\Lambda}_z^{n,k})$ denote the result obtained in the $k$-th inner PG iteration at the $n$-th outer equivalent AO or SPCA iteration. The general PG iteration is given by \cite{Bertsekas1999}
\begin{subequations}\label{gen_PG}
	\begin{align}
	&~~(\bar{P_s}^{n,k+1},\bar{\mathbf{\Lambda}}_z^{n,k+1}) \notag\\
	&=\mathrm{P}_\mathcal{C}(P_s^{n,k}+\mu_{k}\nabla_{P_s}g^{n,k},\mathbf{\Lambda}_z^{n,k}+\mu_{k}\nabla_{\mathbf{\Lambda}_z}g^{n,k}), \label{gen_PG_Pc}\\
	&~~(P_s^{n,k+1},\mathbf{\Lambda}_z^{n,k+1})\notag\\
	&=(P_s^{n,k},\mathbf{\Lambda}_z^{n,k}) +\nu_k(\bar{P_s}^{n,k+1}-P_s^{n,k},\bar{\mathbf{\Lambda}}_z^{n,k+1}-\mathbf{\Lambda}_z^{n,k}),
	\end{align}
\end{subequations}
where $\mu_{k}>0$ and $0<\nu_k\leq1$ are positive step sizes, $\mathrm{P}_\mathcal{C}$ denotes the projection on the feasible set $\mathcal{C}$ in \eqref{constraint}, i.e.,
\begin{align}\label{Pc_PG}
\mathrm{P}_\mathcal{C}(\tilde{P_s},\tilde{\mathbf{\Lambda}}_z)\triangleq \arg\min_{(P_s,\mathbf{\Lambda}_z)\in \mathcal{C}}~(P_s-\tilde{P_s})^2+\|\mathbf{\Lambda}_z-\tilde{\mathbf{\Lambda}}_z\|^2_F,
\end{align}
\begin{figure*}[!t]
	%\vspace*{4pt}
	\begin{subequations}\label{nabla_g}
		\begin{align}
		\nabla_{\mathbf{\Lambda}_z}g^{n,k}=&\beta\mathbf{H}_{tr}^H[(\mathbf{F}^{n,k})^{-1}-\mathbf{S}_0^n]\mathbf{H}_{tr}
		+\mathbf{H}_{te}^H[(\mathbf{R}_e^{n,k})^{-1}-\mathbf{S}_1^n]\mathbf{H}_{te} \nonumber\\
		&~~~~~~~+\sum_{i=1}^L\Big\{\alpha\Tr\left[\left((\mathbf{F}^{n,k})^{-1}-\mathbf{S}_0^n\right)\mathbf{C}_i\right] +\Tr\left[\left((\mathbf{R}_e^{n,k})^{-1}-\mathbf{S}_1^n\right)\mathbf{D}_i\right]\Big\}\mathbf{E}_i,  \\
		\nabla_{P_s}g^{n,k}=&\Tr[(\mathbf{F}^{n,k})^{-1}\mathbf{A}]
		-\Tr(\mathbf{S}_1^n\mathbf{B}).
		\end{align}
	\end{subequations}
	\hrulefill
	\begin{align}\label{nabla_g_def}
	\mathbf{C}_i\triangleq \mathbf{H}_{pr}\mathbf{e}_i\mathbf{e}_i^T\mathbf{H}_{pr}^H,~~
	\mathbf{D}_i\triangleq \mathbf{H}_{pe}\mathbf{e}_i\mathbf{e}_i^T\mathbf{H}_{pe}^H,~~
	\mathbf{E}_i\triangleq \mathbf{H}_{tp}\mathbf{e}_i\mathbf{e}_i^T\mathbf{H}_{tp}^H,~~
	\mathbf{F}^{n,k}\triangleq \mathbf{R}_r^{n,k}+P_s^{n,k} \mathbf{A}.
	\end{align}
	\hrulefill
\end{figure*}
and $(\nabla_{P_s}g^{n,k},\nabla_{\mathbf{\Lambda}_z}g^{n,k})$ denotes the gradient of the function $g(P_s,\mathbf{\Lambda}_z,P_s^{n-1},\mathbf{\Lambda}_z^{n-1})$ in \eqref{objective} w.r.t. $(P_s,\mathbf{\Lambda}_z)$ at the point $(P_s^{n,k},\mathbf{\Lambda}_z^{n,k})$, which is shown in \eqref{nabla_g} at the top of the page where  $\mathbf{C}_i$, $\mathbf{D}_i$, $\mathbf{E}_i$, and $\mathbf{F}^{n,k}$ are defined in \eqref{nabla_g_def} with $\mathbf{S}_0^n$ and $\mathbf{S}_1^n$ defined in \eqref{S01_solu}. %\addtocounter{equation}{2}
To be specific, $\nabla_{\mathbf{\Lambda}_z}g$ here is actually the conjugate derivative of a real function $g$ w.r.t. a Hermitian matrix, which falls into the field of generalized complex-valued matrix derivatives and follows from \cite{Hjorungnes2011}. 
Note that once the initial value for the PG method $(P_s^{n,0},\mathbf{\Lambda}_z^{n,0})$ is chosen to be feasible, i.e. $(P_s^{n,0},\mathbf{\Lambda}_z^{n,0})\in \mathcal{C}$, it is clear that the sequence $\{P_s^{n,k},\mathbf{\Lambda}_z^{n,k}\}$ generated by the PG method is feasible for any fixed $ n $ due to the projection operation in \eqref{gen_PG_Pc} and the condition $0<\nu_k\leq1$. 
By exploiting the structure of the set $\mathcal{C}$, the projection $\mathrm{P}_\mathcal{C}$ in \eqref{Pc_PG} can be formulated in a semi-closed form, which is shown in the following theorem.% \ref{th:Pc}.

\begin{theorem}  \label{th:Pc}
Let $\tilde{\mathbf{\Lambda}}_z=\mathbf{U}\diag\{\tilde{\bm{\eta}}\}\mathbf{U}^H$ be the eigenvalue decomposition of $\tilde{\mathbf{\Lambda}}_z$. Then the optimal solution to problem \eqref{Pc_PG} is given by
\begin{align}\label{Pc_PG_sol}
\mathrm{P}_\mathcal{C}(\tilde{P_s},\tilde{\mathbf{\Lambda}}_z)=(P_s^\star,\mathbf{U}\diag\{\bm{\eta}^\star\}\mathbf{U}^H),
\end{align}
where $P_s^\star$ and $\bm{\eta}^\star$ are unique and take the form \begin{align}\label{Pc_PG_sol_eta}
[P_s^\star,\bm{\eta}^{\star T}]^T =\Big[ [\tilde{P_s},\tilde{\bm{\eta}}^{T}]^T-\lambda^\star\mathbf{1} \Big]^+,
\end{align}
with $\mathbf{1} \triangleq [1,1,\ldots,1]^T$ and the water-filling level $\lambda^\star$ chosen as the minimum nonnegative value such that $P_s^\star+\bm{\eta}^{\star T}\mathbf{1} \leq P$.% (note that if $\lambda^\star>0$, then $(P_s^\star,\bm{\eta}^{\star T})\mathbf{1} = P$).
\end{theorem}
\begin{IEEEproof}
The proof is given in Appendix \ref{app_Pc}.
\end{IEEEproof}

We remark that the water-filling level $\lambda^\star$ in Theorem \ref{th:Pc} can be efficiently obtained by some practical algorithms based on hypothesis testing\cite{Palomar2005}. As for the choosing of the step sizes $\mu_{k}$ and $\nu_k$ in the PG method, several strategies obeying the Armijo rule \cite[Section 2.3.1]{Bertsekas1999} can be exploited. Here, we fix the second step size as $\nu_k=1$, while the backtracking line search\cite{Boyd2004} is adopted to determine the first step size $\mu_{k}$. In this way, the iteration in \eqref{gen_PG} degenerates into
\begin{align}\label{degen_PG}
&(P_s^{n,k+1},\mathbf{\Lambda}_z^{n,k+1}) \notag\\
&~~~~=\mathrm{P}_\mathcal{C}(P_s^{n,k}+\mu_{k}\nabla_{P_s}g^{n,k},\mathbf{\Lambda}_z^{n,k}+\mu_{k}\nabla_{\mathbf{\Lambda}_z}g^{n,k}).
\end{align}
The iteration in \eqref{degen_PG} is guaranteed to converge to the global maximum for the convex optimization problem in  \eqref{general_OP3}\cite{Iusem2003}, and it achieves a good balance between the convergence rate and computational complexity\cite{Jiang2015}. The procedure of backtracking line search for choosing $\mu_{k}$ is listed in Algorithm \ref{alg:bt}. 
%, where the first-order Taylor expansion of the real-valued function $g(\mathbf{\Lambda}_z,\mathbf{\Lambda}_z^{n-1})$ about the complex-valued matrix $\mathbf{\Lambda}_z$ follows from [\ref{ComplexGrad_label}, Theorems 3 and 4]. 
The parameter $\gamma \in (0,1)$, and typical algorithmic parameters are $\mu_0=1$, $\gamma=0.5$, and $\delta=0.1$. Algorithm \ref{alg:bt} is referred to as the Armijo search along the boundary of $\mathcal{C}$\cite{Bertsekas1999}, \cite{Iusem2003}.
\begin{algorithm}
\caption{Backtracking Line Search for Choosing $\mu_{k}$}\label{alg:bt}
\begin{algorithmic}[1]
\REQUIRE $\mu=\mu_0$;
%\ENSURE $\mu_k$;
\WHILE{\TRUE}
\STATE Compute $ (P_s^{n,k+1},\mathbf{\Lambda}_z^{n,k+1}) $ according to \eqref{degen_PG};
\IF{$g^{n,k+1}>g^{n,k}+ \delta \cdot \big\{\Tr[(\nabla_{\mathbf{\Lambda}_z}g^{n,k})^H(\mathbf{\Lambda}_z^{n,k+1}-\mathbf{\Lambda}_z^{n,k})]+\nabla_{P_s}g^{n,k}(P_s^{n,k+1}-P_s^{n,k})\big\}$}
\STATE \textbf{Break};
\ENDIF
\STATE $\mu=\gamma\mu$;
\ENDWHILE
\RETURN $\mu_k=\mu$;
\end{algorithmic}
\end{algorithm}

We summarize our fast algorithm for SRM, combining the PG method for inner convex problem with the formerly proposed outer AO or SPCA iteration, in Algorithm \ref{alg:fast}, where $C_s^n$ and $g^{n,k}$ are used to denote $C_s(P_s^n,\mathbf{\Lambda}_z^n)$ and $g(P_s^{n,k},\mathbf{\Lambda}_z^{n,k},P_s^{n-1},\mathbf{\Lambda}_z^{n-1})$, respectively, for notational convenience. 

The main computational complexity of Algorithm \ref{alg:fast} lies in the  multiplication, inverse, and eigenvalue decomposition of a matrix. 
To facilitate the complexity comparison with \verb|CVX| given in Section \ref{sec_equivalent}, here we %assume the antenna number at the transmitter of the reader  
give the complexity cost of Algorithm \ref{alg:fast} w.r.t. $M$ only, which is $O(JRM^3)$, where $J$ denotes the total iteration number of the outer AO or SPCA method as before and $R$ represents the average iteration number of the inner PG method at each outer iteration.
Recalling the computational complexity of \verb|CVX| given in Section \ref{sec_equivalent}, we see that the improvement in that of the fast PG algorithm is significant. This observation will also be verified by the simulations in Section \ref{sec_sim}.
%Note that in Algorithm \ref{alg:fast}, we adopt a warm-start operation in line \ref{alg_fast_warm}, where the optimal PG solution at the previous outer iteration is used to initialize the PG procedure at the present outer iteration. This warm-start operation can improve convergence rate for the PG procedure in practice\cite{Li2013a}.
%\linespread{0.055}
\begin{algorithm}
\caption{Fast PG Algorithm for SRM}\label{alg:fast}
\begin{algorithmic}[1]
\REQUIRE $P$, $n=1$, $(P_s^0,\mathbf{\Lambda}_z^0) \in \mathcal{C}$, $\epsilon_1>0$, $\epsilon_2>0$;
%\ENSURE $(P_s^\star,\mathbf{\Lambda}_z^\star)$;
\WHILE{$|(C_s^n-C_s^{n-1})/C_s^{n-1}|>\epsilon_1$}
\STATE $k=0,~(P_s^{n,0},\mathbf{\Lambda}_z^{n,0})=(P_s^{n-1},\mathbf{\Lambda}_z^{n-1})$; \label{alg_fast_warm}
\WHILE{$|(g^{n,k}-g^{n,k-1})/g^{n,k-1}|>\epsilon_2$}%The iteration $k$ in $g(\mathbf{\Lambda}_z^{n,k},\mathbf{\Lambda}_z^{n-1})$ not converge for the tolerance $\epsilon_2$}
\STATE Compute $\nabla_{\mathbf{\Lambda}_z}g^{n,k}$ according to \eqref{nabla_g};
\STATE Compute the step size $\mu_{k}$ according to Algorithm \ref{alg:bt};
\STATE Calculate $(\tilde{P_s},\tilde{\mathbf{\Lambda}}_z)=(P_s^{n,k}+\mu_{k}\nabla_{P_s}g^{n,k},\mathbf{\Lambda}_z^{n,k}+\mu_{k}\nabla_{\mathbf{\Lambda}_z}g^{n,k})$;
\STATE Calculate $(P_s^{n,k+1},\mathbf{\Lambda}_z^{n,k+1})=\mathrm{P}_\mathcal{C}(\tilde{P_s},\tilde{\mathbf{\Lambda}}_z)$ according to \eqref{Pc_PG_sol}-\eqref{Pc_PG_sol_eta};
\STATE $k=k+1$;
\ENDWHILE
\STATE $(P_s^n,\mathbf{\Lambda}_z^n)=(P_s^{n,k},\mathbf{\Lambda}_z^{n,k})$;
\STATE $n=n+1$;
\ENDWHILE
\RETURN $(P_s^\star,\mathbf{\Lambda}_z^\star)=(P_s^n,\mathbf{\Lambda}_z^n)$;
\end{algorithmic}
\end{algorithm}

%Note that we have proposed the fast PG algorithm for our general SRM problem in \eqref{general_OP}, 
Note that this fast PG algorithm can be also applied to solve the nullspace SRM problem in \eqref{null_BF} after some modifications. 
%One can easily see that the feasible set in \eqref{null_BF} is actually the same as the set \eqref{constraint} due to the rotational invariance of the trace. 
%The only difference between the two problems is in that we now have constraint \eqref{nullcs}, and the gradient of the function
%$g(P_s,\mathbf{\Lambda}_z(\mathbf{W}),P_s^{n-1},\mathbf{\Lambda}_z(\mathbf{W}^{n-1}))$ w.r.t. $\mathbf{W}$ must be recomputed. 
From the chain rule of the gradient, the gradient of the function $g(P_s,\mathbf{\Lambda}_z(\mathbf{W}),P_s^{n-1},\mathbf{\Lambda}_z(\mathbf{W}^{n-1}))$ w.r.t. $\mathbf{W}$ at the point $\mathbf{W}^{n,k}$ is given by
\begin{align}\label{grad_W}
\nabla_{\mathbf{W}}g^{n,k}=\mathbf{V}^H\nabla_{\mathbf{\Lambda}_z}g^{n,k}\mathbf{V},
\end{align}
where $\mathbf{V}$ is defined in Section \ref{sec_formulation}. With the optimization variable changed to $\mathbf{W}$ and the gradient changed to \eqref{grad_W}, we can still apply Algorithm \ref{alg:fast} to efficiently solve the nullspace SRM problem in \eqref{null_BF}.

\section{Single-Antenna Tag}\label{sec_single_ant}
In the previous sections, we have solved the SRM problem with a multi-antenna tag.
In real applications, due to the resource and cost constrained property of the RFID network, currently a single antenna is usually used at the tag in the market\cite{Zheng2012}. Thus, it is necessary to consider security issues under the scenario where the tag has a single antenna, while the reader and the eavesdropper have multiple antennas. 
It should be noted that previously proposed methods can still be exploited to obtain a \emph{local optimal} solution for this scenario. 
In this section, we focus on finding a low-complexity algorithm which yields the \emph{global optimal} solution to the SRM problem with a single-antenna tag under some practical assumptions. 

When the tag employs a single antenna, 
%after applying some appropriate modifications to the channel notations in Section \ref{sec_model}, 
all the channels form/to the tag reduce to vectors and we redefine $ D_{tp} \triangleq \sqrt{1/M}\mathbf{h}_{tp}^H\mathbf{1}_M $.
%Since the tag now transmits only one information stream, the corresponding achievable rates achieved by the minimum mean square error (MMSE) receiver at the reader and the eavesdropper become
%\begin{equation} \label{C_single_ant}
%\begin{split}
%&C_r^\mathrm{MMSE}=\log_2[1+P_s|D_{tp}|^2\mathbf{h}_{pr}^H(\alpha\mathbf{h}_{pr}\mathbf{h}_{pr}^H(\mathbf{h}_{tp}^H\mathbf{\Lambda}_z\mathbf{h}_{tp})+\beta\mathbf{H}_{tr}\mathbf{\Lambda}_z\mathbf{H}_{tr}^H+\sigma_r^2\mathbf{I})^{-1}\mathbf{h}_{pr}], \\
%&C_e^\mathrm{MMSE}=\log_2[1+P_s|D_{tp}|^2\mathbf{h}_{pe}^H(\mathbf{h}_{pe}\mathbf{h}_{pe}^H(\mathbf{h}_{tp}^H\mathbf{\Lambda}_z\mathbf{h}_{tp})+\mathbf{H}_{te}\mathbf{\Lambda}_z\mathbf{H}_{te}^H+\sigma_e^2\mathbf{I})^{-1}\mathbf{h}_{pe}]. 
%\end{split}
%\end{equation}
%However, it is still difficult to obtain the optimal solution of the similar SRM problem as in \eqref{general_OP} even in this simplified scenario. This is because: i) the problem is still non-convex, and the AN covariance matrix as well as the power allocated to the CW signal needs to be jointly optimized; ii) the terms, such as $ \mathbf{h}_{tp}^H\mathbf{\Lambda}_z\mathbf{h}_{tp} $ and $ \mathbf{H}_{tr}\mathbf{\Lambda}_z\mathbf{H}_{tr}^H $ at the reader, relevant to the optimization variable $ \mathbf{\Lambda}_z $ in \eqref{C_single_ant} cannot be merged in a uniform way, making the optimization hard to perform; and iii) the covariance of the interference and noise term still takes a form of a matrix combined with a matrix inverse operation, which is hard to tackle in the optimization. 
To facilitate analysis and obtain the traceable optimal solution to the SRM problem with a single-antenna tag, here we make the following two assumptions: 
\begin{enumerate}[1)]
	\item The eavesdropper is not aware of the noise injection scheme and thereby simply adopts maximum ratio combining (MRC) to deal with the received signal. %, which is a quite reasonable assumption. 
	Note that when the noise injection scheme is known by the eavesdropper, it may adaptively apply the minimum mean square error (MMSE) receiver to mitigate the jamming from the reader. The SRM problem under this scenario can be similarly tackled by the methods used in the multi-antenna tag case.
	% % also adopted in %TO-DO:add cite;
	\item The reader transmits the AN signal in the nullspace of its self-interference channel, which is a practical assumption mainly in that the reader can equip with only one more antenna for transmitting compared with for receiving. It should be noted that the mathematical model under this assumption is similar to the one under the assumption that the reader can perfectly cancel the AN received from the self-interference channel, i.e., $ \beta=0 $. Thus, for notational simplicity we only consider the latter assumed situation in this section. %Additionally, we will find in the simulation that this nullspace AN scheme can actually achieve the near-optimal performance due to the observation that the self-interference generally dominates in the received signal.
\end{enumerate}
Under the above two assumptions, the achievable rates in \eqref{a_rates} and \eqref{a_rates1} now change to
\begin{align*} %\label{C_single_ant1}
&~~C_r^\mathrm{MMSE,ZF}\\
&=\log_2\big(1+P_s|d_{tp}|^2\mathbf{h}_{pr}^H(\alpha\mathbf{h}_{pr}\mathbf{h}_{pr}^H(\mathbf{h}_{tp}^H\mathbf{\Lambda}_z\mathbf{h}_{tp})+\sigma_r^2\mathbf{I})^{-1}\mathbf{h}_{pr}\big)\\
&\overset{(a)}{=}\log_2\left(1+\frac{P_s|d_{tp}|^2\|\mathbf{h}_{pr}\|^2}{\alpha\|\mathbf{h}_{pr}\|^2(\mathbf{h}_{tp}^H\mathbf{\Lambda}_z\mathbf{h}_{tp})+\sigma_r^2}\right)
\end{align*}
and
\begin{align*}
&~~C_e^\mathrm{MRC}\\
&=\log_2\left(1+\frac{P_s|d_{tp}|^2\|\mathbf{h}_{pe}\|^2}{\|\mathbf{h}_{pe}\|^2(\mathbf{h}_{tp}^H\mathbf{\Lambda}_z\mathbf{h}_{tp})+\frac{\mathbf{h}_{pe}^H\mathbf{H}_{te}\mathbf{\Lambda}_z\mathbf{H}_{te}^H\mathbf{h}_{pe}}{\|\mathbf{h}_{pe}\|^2}+\sigma_e^2}\right), 
\end{align*}
respectively, where $ (a) $ follows from the matrix inverse lemma. 
The SRM problem with a single-antenna tag now becomes
%\begin{subequations}
\begin{align}\label{single}
\max_{P_s,\mathbf{\Lambda}_z}~~&\frac{1+\frac{P_s|d_{tp}|^2\|\mathbf{h}_{pr}\|^2}{\alpha\|\mathbf{h}_{pr}\|^2(\mathbf{h}_{tp}^H\mathbf{\Lambda}_z\mathbf{h}_{tp})+\sigma_r^2}}{1+\frac{P_s|d_{tp}|^2\|\mathbf{h}_{pe}\|^2}{\|\mathbf{h}_{pe}\|^2(\mathbf{h}_{tp}^H\mathbf{\Lambda}_z\mathbf{h}_{tp})+\mathbf{h}_{pe}^H\mathbf{H}_{te}\mathbf{\Lambda}_z\mathbf{H}_{te}^H\mathbf{h}_{pe}/\|\mathbf{h}_{pe}\|^2+\sigma_e^2}} \notag\\
\st ~~&P_s+\Tr(\mathbf{\Lambda}_z ) \leq P,~P_s \geq 0, ~\mathbf{\Lambda}_z\succeq\mathbf{0}.
\end{align}
%\end{subequations}
In the following, we will show that although problem \eqref{single} is still non-convex, its optimal solution is traceable. The basic idea to solve problem \eqref{single} is to reduce the original problem to a single-argument optimization problem, and then the optimal solution can be efficiently obtained by one-dimensional search. Before proceeding, we first give the following lemma about the rank property of the optimal AN covariance $ \mathbf{\Lambda}_z^\star $ for problem \eqref{single}.
\begin{lemma}
	The optimal AN covariance $ \mathbf{\Lambda}_z^\star $ for problem \eqref{single} is rank-one.
\end{lemma}
\begin{IEEEproof}
	To show the optimal $ \mathbf{\Lambda}_z^\star $ is rank-one for problem \eqref{single}, we first let $ \|\mathbf{h}_{pe}\|^2(\mathbf{h}_{tp}^H\mathbf{\Lambda}_z\mathbf{h}_{tp})+\mathbf{h}_{pe}^H\mathbf{H}_{te}\mathbf{\Lambda}_z\mathbf{H}_{te}^H\mathbf{h}_{pe}/\|\mathbf{h}_{pe}\|^2 =s$ be fixed and  the optimal $ \mathbf{\Lambda}_z^\star $ must satisfy%consider the following problem
%	\begin{subequations}
\begin{align}\label{QCQP}
&\mathbf{\Lambda}_z^\star=\arg\min_{\mathbf{\Lambda}_z}~~\mathbf{h}_{tp}^H\mathbf{\Lambda}_z\mathbf{h}_{tp} \notag\\
&\st~~\|\mathbf{h}_{pe}\|^2(\mathbf{h}_{tp}^H\mathbf{\Lambda}_z\mathbf{h}_{tp})+\mathbf{h}_{pe}^H\mathbf{H}_{te}\mathbf{\Lambda}_z\mathbf{H}_{te}^H\mathbf{h}_{pe}/\|\mathbf{h}_{pe}\|^2 = s, \notag\\
&~~~~~~\Tr\{\L \}\leq P,~\L  \succeq\mathbf{0}.
\end{align}
%	\end{subequations}
	To see more clearly, the above problem can be recast as
%	\begin{subequations}
		\begin{align}\label{SDR}
		&\mathbf{\Lambda}_z^\star=\arg\min_{\mathbf{\Lambda}_z}~~\Tr\{\mathbf{C}_1\mathbf{\Lambda}_z\} \notag\\
		&~~~~~~~~\st~~\Tr\{\mathbf{C}_2\mathbf{\Lambda}_z\}= s,~\Tr\{\L \}\leq P,~\L  \succeq\mathbf{0},
		\end{align}
%	\end{subequations}
	where $ \mathbf{C}_1 \triangleq \mathbf{h}_{tp}\mathbf{h}_{tp}^H $ and $ \mathbf{C}_2 \triangleq  \|\mathbf{h}_{pe}\|^2\mathbf{h}_{tp}\mathbf{h}_{tp}^H+\mathbf{H}_{te}^H\mathbf{h}_{pe}\mathbf{h}_{pe}^H\mathbf{H}_{te}/\|\mathbf{h}_{pe}\|^2  $. 
	Problem \eqref{SDR} takes a semidefinite relaxation (SDR) form of a complex-valued homogeneous quadratically constrained quadratic program (QCQP) with two constraints. According to the conclusion in \cite{Luo2010}, the SDR is tight and the solution of problem \eqref{SDR} is rank-one. This completes the proof.
\end{IEEEproof}

In addition, we remark that the optimal solution to problem \eqref{single} must satisfy the total power constraint with equality. This can be  shown by contradictory. Suppose that the optimal solution to problem \eqref{single} is $ (P_s^\star,\L^\star) $ with $P_s^\star+\Tr\{\L^\star\} < P $, then we can construct a new feasible solution $ (P_s^\star,\bar{\mathbf{\Lambda}}_z)  $ such that $P_s^\star+\Tr\{\bar{\mathbf{\Lambda}}_z\} = P $ where $ \bar{\mathbf{\Lambda}}_z=\L^\star+\mathbf{r}\mathbf{r}^H $, $ \mathbf{h}_{tp}^H\mathbf{r}=0 $, and $\mathbf{h}_{pe}^H\mathbf{H}_{te}\mathbf{r} \neq 0  $. It can be easily verified that $ (P_s^\star,\bar{\mathbf{\Lambda}}_z)  $ yields a larger objective value, which is a contradictory.

\begin{figure*}[!t]
	%\vspace*{4pt}
	%	\begin{subequations}
	\begin{align}\label{single1}
	\max_{P_s,\mathbf{v}}~~&\frac{1+\frac{P_s|d_{tp}|^2\|\mathbf{h}_{pr}\|^2}{\alpha\|\mathbf{h}_{pr}\|^2(P-P_s)(\mathbf{v} ^H\mathbf{h}_{tp}\mathbf{h}_{tp}^H\mathbf{v})+\sigma_r^2}}{1+\frac{P_s|d_{tp}|^2\|\mathbf{h}_{pe}\|^2}{\|\mathbf{h}_{pe}\|^2(P-P_s)(\mathbf{v} ^H\mathbf{h}_{tp}\mathbf{h}_{tp}^H\mathbf{v})+(P-P_s)\mathbf{v} ^H \mathbf{H}_{te}^H\mathbf{h}_{pe}\mathbf{h}_{pe}^H\mathbf{H}_{te}\mathbf{v}/\|\mathbf{h}_{pe}\|^2+\sigma_e^2}} 
	~~~~~~~~~\st ~~\|\mathbf{v}\|=1, ~0\leq P_s \leq P.
	\end{align}
	%	\end{subequations}
	\hrulefill
	%	\begin{subequations}
	\setcounter{mytempeqncnt}{\value{equation}}
	\setcounter{equation}{37}
	\begin{align}\label{single2}
	\max_{P_s,t}~~&y(P_s,t) \triangleq \frac{1+\frac{P_s|d_{tp}|^2\|\mathbf{h}_{pr}\|^2}{\alpha\|\mathbf{h}_{pr}\|^2\|\mathbf{h}_{tp}\|^2(P-P_s)t+\sigma_r^2}}{1+\frac{P_s|d_{tp}|^2\|\mathbf{h}_{pe}\|^2}{\|\mathbf{h}_{pe}\|^2\|\mathbf{h}_{tp}\|^2(P-P_s)t+\|\mathbf{H}_{te}^H\mathbf{h}_{pe}\|^2(P-P_s)r(t)/\|\mathbf{h}_{pe}\|^2+\sigma_e^2}} 
	~~~~~~~~~\st ~~0\leq t \leq 1, ~0\leq P_s \leq P.
	\end{align}
	\setcounter{equation}{\value{mytempeqncnt}}
	%	\end{subequations}
	\hrulefill
\end{figure*}

From the above discussions about the properties of the optimal solution to problem \eqref{single}, the AN covariance matrix can be expressed as  $ \L=(P-P_s)\mathbf{v}\mathbf{v} ^H $ where $ \|\mathbf{v}\|=1 $. We can further recast problem \eqref{single} as problem \eqref{single1} shown at the top of the page.
Problem \eqref{single1} is still difficult to handle, to facilitate the further analysis we first consider the optimization of $\mathbf{v}$ in problem \eqref{single1}. Let  $\mathbf{v}^H\mathbf{d}_1\mathbf{d}_1^H\mathbf{v}=t,~0\leq t \leq 1$ where $ \mathbf{d}_1\triangleq \mathbf{h}_{tp}/\|\mathbf{h}_{tp}\| $. By considering the following subproblem
\begin{align}\label{sing_sub}
\begin{split}
&r(t) \triangleq \max_{\mathbf{v}}~~\mathbf{v}^H\mathbf{d}_2\mathbf{d}_2^H\mathbf{v} %\\
~~~~~\st~~\mathbf{v}^H\mathbf{d}_1\mathbf{d}_1^H\mathbf{v}=t, ~\|\mathbf{v}\|=1,
\end{split}
\end{align}
where $ \mathbf{d}_2\triangleq \mathbf{H}_{te}^H\mathbf{h}_{pe}/\|\mathbf{H}_{te}^H\mathbf{h}_{pe}\| $, then problem \eqref{single1} can be reduced to problem \eqref{single2} shown at the top of the  page.\addtocounter{equation}{1}
The following lemma gives the closed-from solution to the subproblem in  \eqref{sing_sub}.
\begin{lemma}
	\cite{Li2011} %Let $ \mathbf{d}_1 $ and $ \mathbf{d}_2 $ be any unit-norm vector. 
	Let $ \phi \in (-\pi,\pi] $ be the argument of $ \mathbf{d}_2^H\mathbf{d}_1,\kappa=|\mathbf{d}_1^H\mathbf{d}_2|\neq 1 $. Then the optimal solution to problem \eqref{sing_sub} and the corresponding objective function value are given by
	\begin{align}
	&\mathbf{v}^\star(t)=\Bigg( \kappa\sqrt{\frac{1-t}{1-\kappa^2}} -\sqrt{t} \Bigg)\mathrm{e}^{\mathrm{i}(\pi-\phi)} \mathbf{d}_1 + \sqrt{\frac{1-t}{1-\kappa^2}} \mathbf{d}_2,\\
	&r(t)=1-\Big(\kappa\sqrt{1-t}-\sqrt{(1-\kappa^2)t}\Big)^2.
	\end{align}
\end{lemma}

Now our aim reduces to solve the two-argument optimization problem in \eqref{single2}. To solve problem \eqref{single2}, our method is to first find the optimal $ P_s^\star(t) $ for any given $ t $ and then perform the one-dimensional search w.r.t. $ t $. To facilitate analysis, for any given $t$ we recast the objective function in problem \eqref{single2} w.r.t. $P_s$ as
\begin{align}\label{y_Ps}
y^t(P_s)=\frac{1+\frac{a P_s}{1+b P_s}}{1+\frac{c P_s}{1+d P_s}},
\end{align}
where
\begin{align*}
&a \triangleq |d_{tp}|^2\|\mathbf{h}_{pr}\|^2/(\sigma_r^2+Pl_1),
~b  \triangleq -l_1/(\sigma_r^2+Pl_1), \notag\\
&c   \triangleq |d_{tp}|^2\|\mathbf{h}_{pe}\|^2/(\sigma_e^2+Pl_2),
~d \triangleq -l_2/(\sigma_e^2+Pl_2),\\
&l_1 \triangleq \alpha\|\mathbf{h}_{pr}\|^2\|\mathbf{h}_{tp}\|^2t,
\notag\\
&l_2 \triangleq \|\mathbf{h}_{pe}\|^2\|\mathbf{h}_{tp}\|^2t+\|\mathbf{H}_{te}^H\mathbf{h}_{pe}\|^2r(t)/\|\mathbf{h}_{pe}\|^2.
\end{align*}
Let the first order derivative of $ y^t(P_s) $ in \eqref{y_Ps} be zero, we obtain
\begin{align}\label{Dy_Ps}
(a d^2+a c d-a b c-b^2 c)P_s^2+2(ad-bc)P_s+(a-c)=0.
\end{align}
The solution to the above quadratic equation is given by
\begin{align}\label{Dy_Ps_sol}
P^t_{s,1(2)}=\frac{ad-b c\pm\sqrt{a c (b-d) (a+b-c-d)}}{b^2 c+a (b c-d (c+d))},
\end{align}
and the optimal power allocated to the CW signal at the reader for any given $t$ becomes
\begin{equation}\label{Ps_star}
P_s^\star(t) = \arg\max_{P_s}~ y(P_s,t), ~P_s \in \{P^t_{s,1},P^t_{s,2},P\}\cap [0,P],
\end{equation} 
which can be easily computed. Substituting \eqref{Ps_star} into problem \eqref{single2} yields a single-argument optimization problem,
%\begin{subequations}\label{single3}
%	\begin{align}
%	\max_{t}~~&\frac{1+\frac{P_s^\star(t)|d_{tp}|^2\|\mathbf{h}_{pr}\|^2}{\alpha\|\mathbf{h}_{pr}\|^2\|\mathbf{h}_{tp}\|^2(P-P_s^\star(t))t+\sigma_r^2}}{1+\frac{P_s^\star(t)|d_{tp}|^2\|\mathbf{h}_{pe}\|^2}{\|\mathbf{h}_{pe}\|^2\|\mathbf{h}_{tp}\|^2(P-P_s^\star(t))t+\|\mathbf{H}_{te}^H\mathbf{h}_{pe}\|^2(P-P_s^\star(t))r(t)/\|\mathbf{h}_{pe}\|^2+\sigma_e^2}} \label{single3_obj}\\
%	\st ~~&0\leq t \leq 1,
%	\end{align}
%\end{subequations}%some one-dimension optimization methods.
which can be handled by searching $ t $ in the interval $ [0,1] $. Once we have obtained the optimal $ t^\star $, the optimal power allocated to the CW signal and the optimal AN covariance matrix are given by  $P_s^\star=P_s^\star(t^\star)$ and $\L^\star=(P-P_s^\star)\mathbf{v}^\star(t^\star)\mathbf{v}^\star(t^\star) ^H $, respectively.

As in Section \ref{sec_formulation}, here we also consider a suboptimal nullspace AN scheme. More specifically, we can restrict the transmitted AN from the reader to lie in the nullspace of the reader-tag channel $ \mathbf{h}_{tp} $.  
Under this condition, we directly have $ t=0 $ and the only remaining thing is to obtain $ P_s^\star(0) $ by solving the  low-complexity problem in \eqref{Ps_star}. Thus, the computational complexity can be vastly reduced. 
Note that when the total available power or the number of antennas  at the transmitter of the reader is large, this nullspace AN constraint incurs a loss of only one degree of freedom which is  negligible compared with the large number of antennas. 
So from this view, this scheme is practical and beneficial for resource-constrained RFID devices and it achieves a good trade-off between secrecy performance and computational complexity. 
This observation will be verified by the simulation in the next section.

\section{Simulation Results}\label{sec_sim}
This section presents some numerical results to evaluate the secrecy rate performance of the proposed noise-injection precoding schemes as well as their computational efficiency. 
In the simulations, the self-interference channel at the reader is generated as $\mathbf{H}_{tr}=\tilde{\mathbf{H}}_{tr}$, and the other channels are assumed to undergo a path loss combined with a small-scale fading, namely $\mathbf{H}_k=d_k^{-\gamma/2}\tilde{\mathbf{H}}_k$, $k\in \{tp,te,pe,pr\}$, where $d_k$ is the distance between two nodes, $\gamma$ is the path loss exponent, and each element of $\tilde{\mathbf{H}}_j$, $j\in \{tr,tp,te,pe,pr\}$, is an independent and identically distributed (i.i.d.) complex Gaussian random variable with zero mean and unit variance. 
Note that the statistic distribution of the self-interference channel at the reader has not been well understood yet \cite{Alexandris2014}. Thus, for simplicity we adopt this distribution here as in \cite{Zheng2015}.

%In our simulations, we test the performance of the general AN design, and compare it with the suboptimal but low-complexity nullspace AN schemes as well as the scheme without AN.
%For convenience, we refer to the nullspace AN design where 
The simulation settings are as follows, unless otherwise specified: The antenna numbers at the receiver and the transmitter of the reader, the tag, and the eavesdropper are $N=2$, $M=3$, $L=2$, and $K=3$, respectively. 
%The step size of one-dimensional search in solving problem \eqref{single3} in the single-antenna tag case is set as $ 0.01 $.
%As for a RFID reader, the maximum transmission power is typically 30 dBm or 1 Watt\cite{Dobkin2007}. Here we assume that the typical 
The typical transmit power at the reader $P=10$ dBm, and the AWGN power at the reader and the eavesdropper $\sigma_e^2=\sigma_r^2=-20$ dBm. The path loss exponent is set as $\gamma=2$, and we set typical distances between two nodes as $d_k=2~\mathrm{m},~k\in \{tp,te,pe,pr\}$. 
%, and typical reader-tag distances of $d_{RT} = 2$ m and $d_{RT} = 4$ m are assumed\cite{Dobkin2007}. %All channels are generated follow the setting in Section \ref{sec_model}. 
We initialize our algorithms with the initial power allocation $\rho^0=1,~P_s^0=\rho^0 P, ~\mathbf{\Lambda}_z^0=(1-\rho^0)(P/M)\mathbf{I}$ and set with the termination parameters $\epsilon_1=10^{-3}$ and $\epsilon_2=10^{-5}$. 
In addition, to obtain a larger secrecy rate in the general AN design, we use the solution obtained by the nullspace AN scheme as the initial parameter. %  in solving the general SRM problem in \eqref{general_OP}.
Note that under the above antenna number setting the reader can perform two nullspace AN schemes, namely, the no backscattered AN (NBS-AN) scheme and the no self-interference AN (NSI-AN) scheme where the transmitted AN lies in the nullspace of the reader-tag channel and the self-interference channel, respectively.
All results to be shown are averaged over 1000 randomly generated channel realizations.

\begin{figure}[!t]
\centering
\subfloat[]{\includegraphics[width=3.0in]{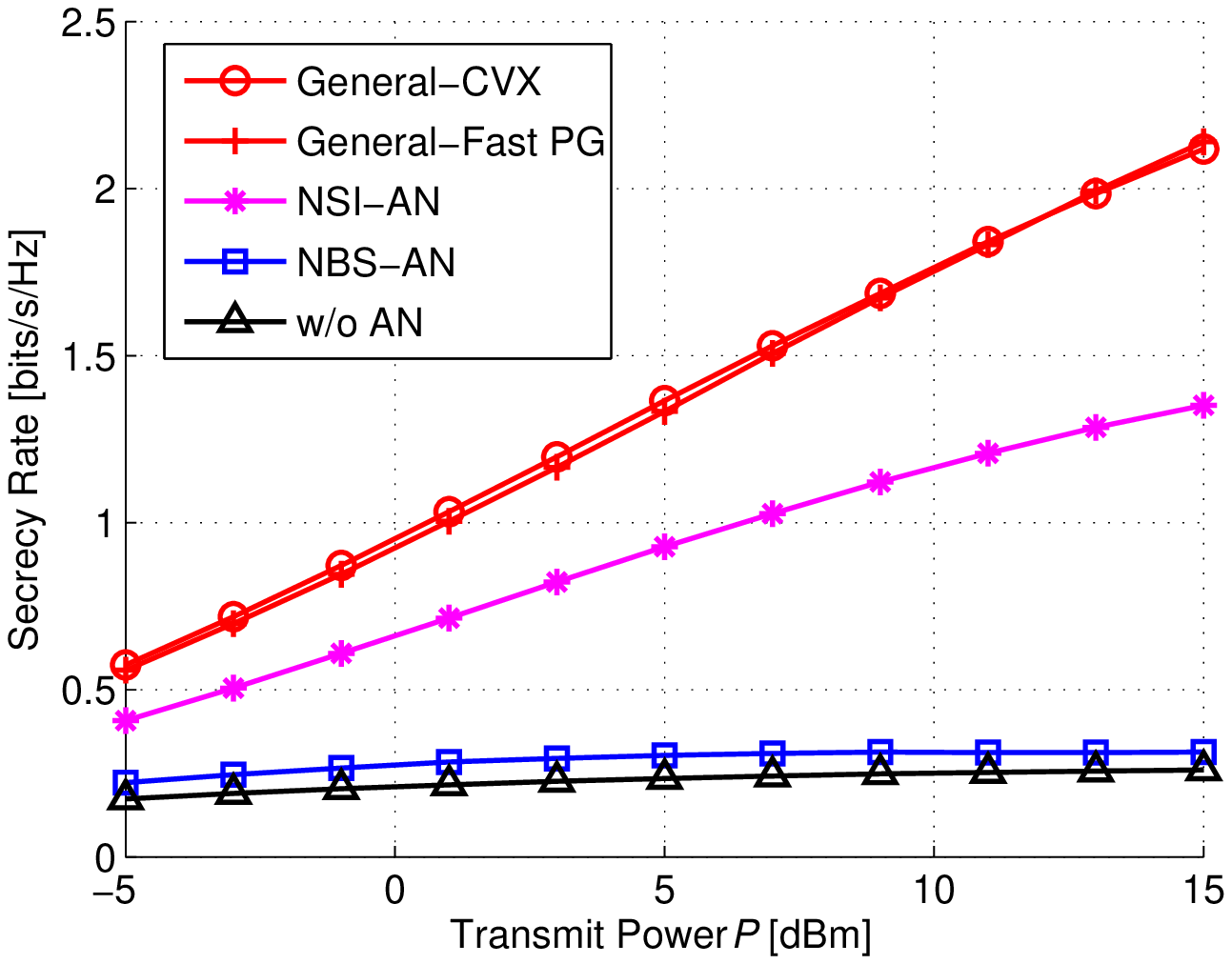}
\label{m_Cs_P}}
\hfil
\subfloat[]{\includegraphics[width=3.0in]{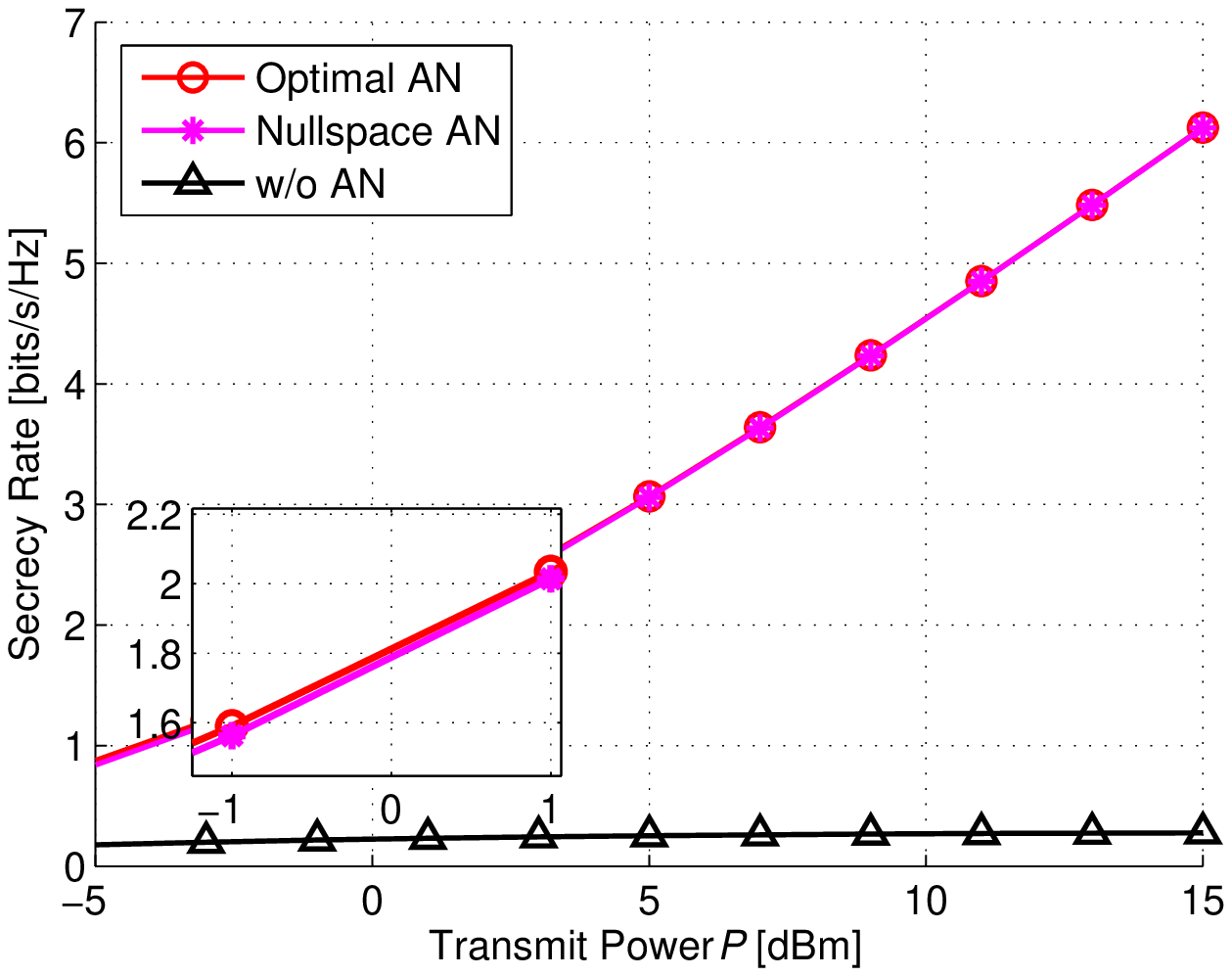}
\label{s_Cs_P}}
\caption{The secrecy rates $C_s$ achieved by different schemes versus the total transmit power of the reader $P$ under (a) multi-antenna and (b) single-antenna tag cases. For both the two cases, the attenuation factors of the backscattered AN are the same $\alpha=0.6$. For the multi-antenna tag case, the attenuation factor of the self-interference $\beta=0.3$, and $\beta=0$ for the single-antenna tag case. %, the AWGN power at the reader and the eavesdropper $\sigma_r^2=\sigma_e^2=-50$ dBm, the distances between any two nodes are the same $d=2$ m.
	}
\label{fig_Cs_P}\vspace{-3mm}
\end{figure}

\begin{table}[!t]
	\caption{Average Running Time (in secs.) Versus Transmit Power}
%	 under Multi-Antenna Tag Case
	\centering
	\setlength{\tabcolsep}{0.8pt}
%	\begin{tabular}{|c|c|c|c|c|c|}
%		\hline
%		\multirow{2}{*}{Method} & \multicolumn{5}{c|}{Transmit Power (dBm)} \\
%		\cline{2-6}
%		& -3 & 1 & 5 & 9 & 13 \\
%		\hline
%		General-\verb"CVX" &  11.8927  &  12.1476  &  12.3213  &  12.5497  &  12.4153 \\
%		\textbf{General-Fast PG} & \textbf{0.2266}  &  \textbf{0.2725}  &  \textbf{0.3118}  &  \textbf{0.3787}  &  \textbf{0.4902} \\
%		\hline
%	\end{tabular}
\begin{tabular}{|c|c|c|c|c|c|c|}
\hline
\multicolumn{2}{|c|}{\multirow{2}{*}{Method}} & \multicolumn{5}{c|}{Transmit Power (dBm)} \\
\cline{3-7}
\multicolumn{2}{|c|}{} & -3 & 1 & 5 & 9 & 13 \\
\hline
\multirow{2}{*}{\tabincell{c}{Multi-Antenna \\Tag Case}} & General-\verb"CVX" &  11.8927  &  12.1476  &  12.3213  &  12.5497  &  12.4153 \\
&\textbf{General-Fast PG} & \textbf{0.2266}  &  \textbf{0.2725}  &  \textbf{0.3118}  &  \textbf{0.3787}  &  \textbf{0.4902} \\
%&$ \mathrm{Null}(\mathbf{H}_{tr})  $ AN&  0.2793  &  0.2881  &  0.2651  &  0.2610  &  0.3256 \\
%&$ \mathrm{Null}(\mathbf{H}_{tp}^H)  $ AN&  0.2196  &  0.2415  &  0.1822  &  0.1328  &  0.0819 \\
\hline
\multirow{2}{*}{\tabincell{c}{Single-Antenna \\Tag Case}}  & Optimal AN &  0.0164  &  0.0164  &  0.0164  &  0.0164  &  0.0164 \\
& \textbf{Nullspace AN} &  \textbf{0.0006} &  \textbf{0.0006}  &  \textbf{0.0006}  &  \textbf{0.0006}  &  \textbf{0.0006} \\
\hline
\end{tabular}
	\label{tab_runtime}
	\vspace{-3mm}
\end{table}
Fig. \ref{fig_Cs_P} shows the secrecy rates achieved by different schemes versus the transmit power $P$ under both the multi-antenna tag case (cf. Fig. \ref{fig_Cs_P}(a)) and the single-antenna tag case (cf. Fig. \ref{fig_Cs_P}(b)). 
In the multi-antenna tag case, we evaluate the performance of the general AN design obtained by the proposed fast PG algorithm in Algorithm \ref{alg:fast} (labeled as ``General-Fast PG''), and compare it with the general AN design obtained by \verb"CVX" (labeled as ``General-CVX''), the two nullspace AN schemes%(labeled as ``NSI-AN'' and ``NBS-AN'')
, and the scheme without AN. %(labeled as ``w/o AN'').
%The figure clearly demonstrates that the secrecy rate increases monotonically with an increase in transmit power. 
It can be observed from Fig. \ref{fig_Cs_P}(a) that the advantage of the proposed noise-injection scheme is significant compared with the scheme without AN especially when $ P $ is large.
As shown in Fig. \ref{fig_Cs_P}(a), the general AN design outperforms the two nullspace AN schemes as it is free from the nullspace AN constraint. One can also see that the proposed fast PG algorithm achieves almost the same secrecy rate as %the general-purpose optimization package
\verb"CVX" does.
In addition, to illustrate the relative computing efficiency of the proposed algorithms for SRM, we present the corresponding average running time in Table \ref{tab_runtime}\footnote{
	The average running time listed in Table I is obtained by a desktop with MATLAB as the simulation tool. This result is only for the purpose of relative comparison between different algorithms, and the measurement of the practical running time on a typical RFID device is out of the scope of this paper.}. 
One can see that the fast PG algorithm is much faster than \verb"CVX". 
In the single-antenna tag case, we compare the performance of the optimal AN design with the nullspace AN precoding. %(labeled as ``$ \mathrm{Null}(\mathbf{h}_{tp}^H) $ AN''). 
It can be observed from Fig. \ref{fig_Cs_P}(b) that the secrecy rate achieved by the nullspace AN precoding is very close to the optimal one. % especially when $ P $ is in moderate and large ranges.
Moreover, through Table \ref{tab_runtime} we see that the nullspace AN precoding enjoys a much lower computational complexity compared with the optimal AN design.
The nullspace AN precoding here achieves a good trade-off between secrecy performance and computational complexity.
These low-complexity algorithms are especially beneficial to resource-constrained RFID devices.

%\begin{figure}[!t]
%	\centering
%	\includegraphics[width=3.0in]{fig_conver}
%	\caption{The convergence of the fast PG algorithm under different initial power allocations $ \rho^0 $ in a channel realization. The total transmit power at the reader $P=10$ dBm, and the attenuation factors of the backscattered AN and self-interference are the same $ \alpha=\beta=0.3 $.}
%	\label{fig_conver} \vspace{-3mm}
%\end{figure}
%Fig. \ref{fig_conver} shows the convergence of the proposed fast PG algorithm in Algorithm \ref{alg:fast} under different initial power allocations $ \rho^0 $ in a channel realization. 
%It can be seen the sequence of the secrecy rates generated by the outer equivalent AO or SPCA method is non-descending, which is expected and guaranteed by Theorem \ref{th_converge}.
%Moreover, the fast PG algorithm converges very quickly (within 15 outer iterations from Fig. \ref{fig_conver}) regardless of different initial power allocations. 
%This quick convergence property provides the fast PG algorithm with much practical significance in the RFID network.

%This implies that the convergence of the proposed fast PG algorithm is insensitive to the initial parameter 
%Note that the  secrecy rates obtained by the inner PG iteration between any two adjacent outer iterates, i.e. markers in Fig. \ref{fig_conver}, may not preserve the non-descending property, because the PG iteration is carried out with the approximated secrecy rate $ g(P_s,\mathbf{\Lambda}_z,P_s^{n-1},\mathbf{\Lambda}_z^{n-1})  $ as its objective function. In contrast to the secrecy rates obtained by the PG iteration, 

\begin{figure}[t]
	\centering
	\subfloat[]{\includegraphics[width=3.0in]{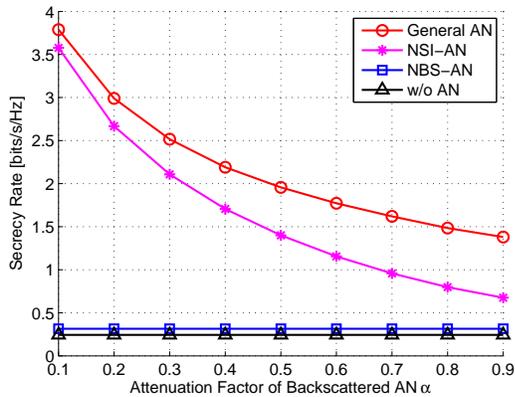}	\label{fig_Cs_alpha}}
	\hfil
	\subfloat[]{\includegraphics[width=3.0in]{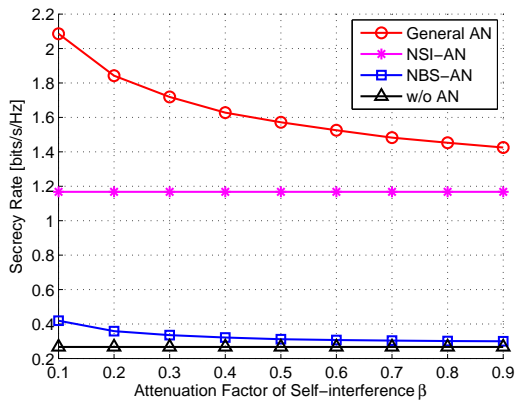}  \label{fig_Cs_beta}}
	\caption{(a) The secrecy rates achieved by different schemes versus the attenuation factor of the backscattered AN $\alpha$ where the attenuation factor of the self-interference is fixed as $\beta=0.3$, and (b) the secrecy rates achieved by different schemes versus the attenuation factor of the self-interference $\beta$ where the attenuation factor of the backscattered AN is fixed as $\alpha=0.6$. Both figures are under the multi-antenna tag case. }
	\label{fig_antenuation}
 \vspace{-3mm}
\end{figure}
Fig. \ref{fig_antenuation}(a) plots the secrecy rates achieved by different schemes versus the attenuation factor of the backscattered AN $\alpha$ under the multi-antenna tag case. From the figure, we see that the secrecy rates decreases as $\alpha$ increases.
Note that the secrecy rate achieved by the NBS-AN scheme remains constant. This is because the injected noise is fully nulled out at the tag in this scheme. 
From Fig. \ref{fig_antenuation}(a), the secrecy rate achieved by the general AN design drops down significantly and approaches to the one obtained by the NBS-AN scheme when $ \alpha $ becomes larger. This is because the backscattered AN received by the reader cannot be well attenuated at this time. 
%Again, Fig. \ref{fig_antenuation}(a) clearly shows the significant benefit of the noise-injection precoding with AN over the scheme without AN. 
Another interesting observation is that the performance of the NSI-AN scheme is close to the general AN design when $\alpha$ is small, and the performance gap of the two schemes becomes larger as $\alpha$ increases. This is because the NSI-AN scheme is close to the optimal scheme only when the self-interference dominates, namely $\alpha$ is small.

Fig. \ref{fig_antenuation}(b) depicts the secrecy rates achieved by different schemes against the attenuation factor of the self-interference $\beta$ under the multi-antenna tag case. We can clearly see that the secrecy rates decrease with an increase in $\beta$. 
Note that the secrecy rate achieved by the NSI-AN scheme remains constant, because the self-interference is fully nulled out at the reader in this scheme. 
It can be seen from Fig. \ref{fig_antenuation} that no matter how $ \alpha $ or $ \beta $ changes the NSI-AN scheme is always superior to the NBS-AN scheme. This is because the received backscattered AN at the reader goes through both the reader-tag and the tag-reader channels and thus experience a double path loss compared with the received AN from the self-interference channel. Hence, the received AN due to self-interference generally dominates in the received signal at the reader, and thus 
the secrecy rate achieved by the NSI-AN scheme is much closer to that achieved by the general AN design. Similarly, this nullspace scheme can serve as a low-complexity method which achieves a good trade-off between secrecy performance and computational complexity in the RFID system.

\begin{figure}[t]
	\centering
	\subfloat[]{\label{fig_ant_M}
		\includegraphics[width=3.0in]{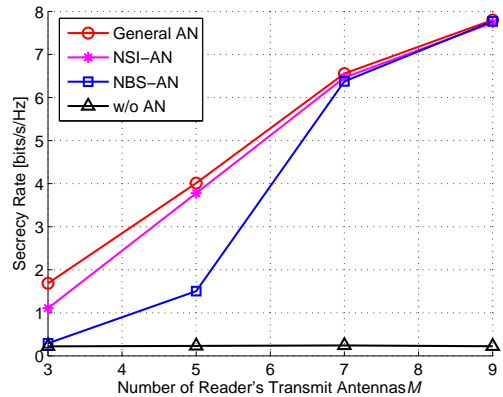}
	}
	\hfil
	\subfloat[]{\label{fig_ant_K}
		\includegraphics[width=3.0in]{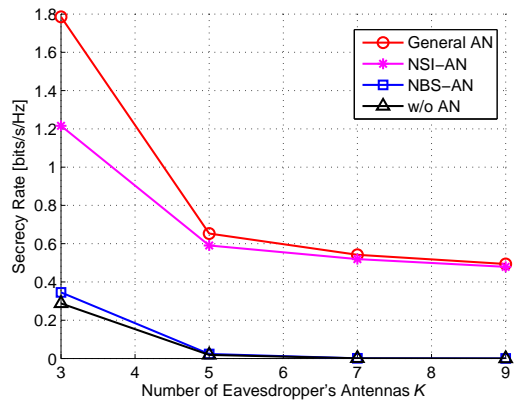}   
	}
	\caption{(a) The secrecy rates achieved by different schemes versus the number of the reader's transmit antennas $M$, and (b) the secrecy rates achieved by different schemes versus the number of the eavesdropper's antennas $K$. The attenuation factors of the backscattered AN and self-interference are $\alpha=0.6$ and $\beta=0.3$, respectively. Both figures are under the multi-antenna tag case.}
	\label{fig_Cs_ant} \vspace{-3mm}
\end{figure}
Fig. \ref{fig_Cs_ant} shows the secrecy rates achieved by different schemes when we increase the number of the reader's transmit antennas $M$ (cf. Fig. \ref{fig_Cs_ant}(a)) or the eavesdropper's antennas $K$ (cf. Fig. \ref{fig_Cs_ant}(b)) under the multi-antenna tag case. 
%Again, we can see that the general AN design optimizing the full AN covariance matrix yields the best performance among all the methods, and 
Again, we can see that the superiority of noise-injection schemes on SRM is significant. 
Note that from Fig. \ref{fig_Cs_ant}(b), the NBS-AN scheme and the scheme without AN cannot even achieve a positive secrecy rate when the number of the eavesdropper's antennas is larger than six. %satisfies $K>6$. 
As shown in Fig. \ref{fig_Cs_ant}(a), the two nullspace AN schemes achieve almost the same secrecy rate as the general AN design does, when the number of the reader's transmit antennas satisfies $M > 7$. This is not surprising because a large number of transmit antennas brings abundant spatial degrees of freedom and the performance loss incurred by the nullspace AN constraint is negligible. Under this situation, it is beneficial for the reader to perform the nullspace AN schemes to reduce the computational complexity and obtain the near-optimal performance.

\begin{figure}[t]
	\centering
	\includegraphics[width=3.0in]{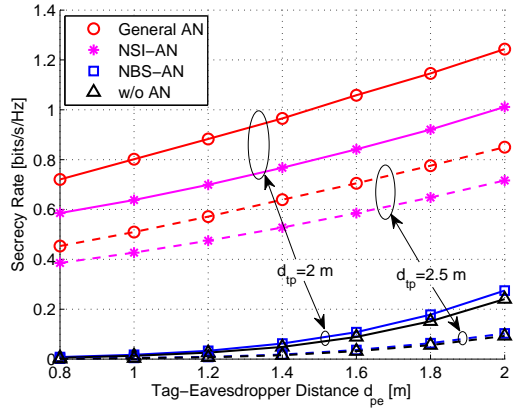}
	\caption{The secrecy rates achieved by different schemes versus the tag-eavesdropper distance $d_{pe}$ under the multi-antenna tag case. The reader-tag distance $d_{tp}$ is set to either 2 m (indicated by the solid line) or 2.5 m (indicated by the dashed line). The reader, the tag, and the eavesdropper are located on a straight line in this order. The attenuation factors of the backscattered AN and self-interference are $\alpha=0.6$ and $\beta=0.3$, respectively.}
	\label{fig_Cs_dte} \vspace{-3mm}
\end{figure}
We then study the impact of the eavesdropper's location on the secrecy rate performance. For simplicity, we assume that the reader, the tag, and the eavesdropper are located on a straight line in this order. %Since in this setting the eavesdropper can be very close to the tag, while the distance from the reader to the tag remains fixed. Thus, this assumption is actually a worst case.
Fig. \ref{fig_Cs_dte} shows the secrecy rates achieved by different schemes versus the tag-eavesdropper distance $d_{pe}$ ranging from 0.8 m to 2 m under the multi-antenna tag case, where the reader-tag distance $d_{tp}$ is set to either 2 m (indicated by the solid line) or 2.5 m (indicated by the dashed line). 
From Fig. \ref{fig_Cs_dte}, we see that the secrecy rate strongly depends on the distances between the nodes due to the power-law decay of the path loss. Indeed, the secrecy rate under the small reader-tag distance is larger than that under the large one. Moreover, the secrecy rate increases significantly as the tag-eavesdropper distance increases.
% Again, Fig. \ref{fig_Cs_dte} demonstrates the improvement, achieved by the noise-injection method, on the secrecy rate especially when the tag-eavesdropper distance is large. 
%Note that the scheme without AN cannot even achieve a positive secrecy rate when the tag-eavesdropper distance $d_{TE}< 0.6$ m from Fig. \ref{fig_Cs_dte}.
% Nevertheless, the optimal scheme outperforms the nullspace AN scheme especially when the reader-tag distance is large or the tag-eavesdropper distance is small. 
In particular, when $d_{tp}=2$ m and the eavesdropper is very close to the tag (e.g. $d_{pe}=0.8$ m), from Fig. \ref{fig_Cs_dte} the general AN design can achieve a positive secrecy rate larger than 0.7 bits/s/Hz, % , 
while the scheme without AN cannot even achieve a positive secrecy rate. This implies that the proposed noise-injection scheme and the optimization of the AN covariance matrix can give the secrecy rate a sharp boost when the eavesdropper is very close to the tag.
%It should be noted that the optimal result here, obtained under the MIMO system setting, is different from the result obtained under a SISO one in \cite{Saad2014} where the secrecy rate sharply decrease when $d_{TE}$ gets very small. This is because the diversity of the MIMO channels, from the reader to the tag and the eavesdropper in this paper, can be further exploited to improve the secrecy rate though the tag-eavesdropper distance is small. This actually reflects a compromise between the performance and the complexity that a better secrecy rate is only achieved by the scheme with a more computational complexity.

%We remark that, in the RFID scenario, the worst tag-eavesdropper distance in practice is actually determined by the actual application scenario considered. For example, if a actual physical protection can prevent the eavesdropper from being closer than a certain distance $d_{TE}^{(\min)}$, then $d_{TE}^{(\min)}$ represents the worst tag-eavesdropper distance in this scenario. We can use this worst tag-eavesdropper distance to obtain the least transmit power needed at the reader by a bisection search (cf. Algorithm \ref{alg:1}).

\section{Conclusion} \label{sec_conclusion}
In this paper, we have studied security issues in a MIMO RFID backscatter system from the perspective of PLS. 
First, %to strengthen the system security, 
we have proposed a noise-injection precoding scheme. 
%Then, we have formulated the SRM problem and subsequently by exploiting the AO and SPCA methods, respectively, we have changed the non-convex SRM problem to a sequence of convex ones.
Then, we have changed the non-convex SRM problem to a sequence of convex ones by exploiting the AO and SPCA methods, respectively.
Interestingly, we have shown a fact that the two methods are actually equivalent for our SRM problem. 
%In addition, the equivalent method is guaranteed to converge to a KKT point. 
Moreover, to facilitate the implementation for resource-constrained RFID devices, a fast algorithm based on the PG method has been proposed. % to efficiently solve the SRM problem. 
As a complement, 
we have studied the single-antenna tag case and derived an algorithm yielding the global optimal solution.
%the case where the tag equips with a single antenna has been studied, wherein the global optimal solution is obtained. 
Numerical results show the superior secrecy rate performance of the proposed noise-injection precoding schemes and the low computational complexity of the proposed algorithms.
Furthermore, the proposed nullspace schemes can achieve a good balance between secrecy performance and computational complexity for the resource-constrained RFID system.

%For future work, one practical aspect is to investigate how to obtain the optimal signal design when the channel state information is unknown or partly known. Other extensions can address a variety of issues such as more efficient, energy saving, and low complexity design under various scenarios and backscatter radio propagation environments.

\begin{appendices}
\section{Proof of Theorem \ref{th_converge}}\label{app_converge}
Here we only show the convergence of the SPCA method for our SRM problem, and the same convergence result holds for the AO method due to the equivalence of the two methods in Section \ref{sec_equivalent}.
We divide the proof into two steps: First, we show that the SPCA method produces non-descending achievable secrecy rates and converges to a limit point; Second, we further show that the method converges to a KKT point of the SRM problem in \eqref{general_OP}.

%To show the convergence of Algorithm \ref{alg:SPCA}, we first show that the optimization problem in \eqref{general_OP2} produces a non-descending achievable secrecy rate at each iteration. 
At the $n$-th iteration of the SPCA method, let $(P_s^n,\mathbf{\Lambda}_z^n)$ be the optimal solution to problem \eqref{general_OP2}, and it is straightforward to see that the optimal solution $(P_s^{n-1},\mathbf{\Lambda}_z^{n-1})$ at the $(n-1)$-th iteration is only a feasible solution at the $n$-th iteration of the SPCA method. 
%Thus, the optimal objective function value at the $n$-th SPCA iteration satisfies
Thus, we have
\begin{align}\label{conver0}
&~~~~f_0(P_s^n,\mathbf{\Lambda}_z^n)-f_1(P_s^n,\mathbf{\Lambda}_z^n,P_s^{n-1},\mathbf{\Lambda}_z^{n-1}) \notag\\
&~~~~~~~~-f_2(P_s^n,\mathbf{\Lambda}_z^n,P_s^{n-1},\mathbf{\Lambda}_z^{n-1}) \nonumber\\
&\geq f_0(P_s^{n-1},\mathbf{\Lambda}_z^{n-1})-f_1(P_s^{n-1},\mathbf{\Lambda}_z^{n-1},P_s^{n-1},\mathbf{\Lambda}_z^{n-1}) \notag\\
&~~~~~~~~-f_2(P_s^{n-1},\mathbf{\Lambda}_z^{n-1},P_s^{n-1},\mathbf{\Lambda}_z^{n-1}) \nonumber\\
%&=f_0(P_s^{n-1},\mathbf{\Lambda}_z^{n-1})-\ln\det(\mathbf{R}_r^{n-1})
%-\ln\det(\mathbf{R}_e^{n-1} +P_s^{n-1} \mathbf{B}) \nonumber\\
&=C_s(P_s^{n-1},\mathbf{\Lambda}_z^{n-1}).
\end{align}
On the other hand, from \eqref{f_approxi} we have
\begin{align}\label{conver1}
&f_0(P_s^n,\mathbf{\Lambda}_z^n)-f_1(P_s^n,\mathbf{\Lambda}_z^n,P_s^{n-1},\mathbf{\Lambda}_z^{n-1}) \notag\\
&~~~~~~~~-f_2(P_s^n,\mathbf{\Lambda}_z^n,P_s^{n-1},\mathbf{\Lambda}_z^{n-1})  %\nonumber\\
%&\leq f_0(P_s^{n},\mathbf{\Lambda}_z^{n})-\ln\det(\mathbf{R}_r^{n})
%-\ln\det(\mathbf{R}_e^{n} +P_s^{n} \mathbf{B}) \nonumber\\
%&=
\leq C_s(P_s^{n},\mathbf{\Lambda}_z^{n}).
\end{align}
Combining \eqref{conver0} with \eqref{conver1} leads to $C_s(P_s^{n},\mathbf{\Lambda}_z^{n})\geq C_s(P_s^{n-1},\mathbf{\Lambda}_z^{n-1})$, which completes the proof for the non-descending property of the achievable secrecy rates produced by the SPCA method.
Note that the achievable secrecy rate $C_s$ is up-bounded for any given transmit power $P$. 
%Thus from the Bolzano-Weierstrass theorem, 
Thereby we conclude that the proposed SPCA method converges to a limit point.

The KKT-point convergence result can be straightforwardly verified from  \cite[Theorem 1]{Marks1978}. From \cite[Step 1]{Marks1978}, we know that the first-order Taylor's series approximation in \eqref{log_es} satisfies the constraints for ensuring the SPCA method. Then from \cite[Theorem 1]{Marks1978}, the convergence to a KKT point of the original problem is guaranteed. 
This completes the proof.

\section{Proof of Theorem \ref{th:Pc}}\label{app_Pc}
To fulfill the proof, we first illustrate the original projection problem in \eqref{Pc_PG} can be reduced to a simplex projection problem given by
\begin{align}\label{Pc_PG_simplex}
(P_s^\star,\bm{\eta}^{\star})=&\arg\min_{P_s,\bm{\eta}}~(P_s-\tilde{P_s})^2+\|\bm{\eta}-\tilde{\bm{\eta}}\|^2 \notag\\
&~~~~~~\st~~P_s\geq 0,~\bm{\eta}\geq\mathbf{0},~P_s+\bm{\eta}^{T}\mathbf{1} \leq P.
\end{align}

By unitary invariance of the Frobenius norm, problem \eqref{Pc_PG} is equivalent to
\begin{align}
&\min_{(P_s,\mathbf{\Lambda}_z)\in \mathcal{C}}~(P_s-\tilde{P_s})^2+\|\mathbf{\Lambda}_z-\mathbf{U}\diag\{\tilde{\bm{\eta}}\}\mathbf{U}^H\|^2_F  \nonumber\\
\Longleftrightarrow&\min_{(P_s,\mathbf{\Lambda}_z)\in \mathcal{C}}~(P_s-\tilde{P_s})^2+\|\mathbf{U}^H\mathbf{\Lambda}_z\mathbf{U}-\diag\{\tilde{\bm{\eta}}\}\|^2_F \nonumber\\
%\mathop{\Longleftrightarrow}\limits^{(a)}
\Longleftrightarrow&\min_{(P_s,\hat{\mathbf{\Lambda}}_z)\in \mathcal{C}}~(P_s-\tilde{P_s})^2+\|\hat{\mathbf{\Lambda}}_z-\diag\{\tilde{\bm{\eta}}\}\|^2_F, \label{Pc_PG_diag}
\end{align}
where $\hat{\mathbf{\Lambda}}_z \triangleq \mathbf{U}^H\mathbf{\Lambda}_z\mathbf{U}$. It can be easily seen by contradiction that the optimal solution $\hat{\mathbf{\Lambda}}_z^\star$ to problem \eqref{Pc_PG_diag} must take the form of a diagonal matrix.
%Suppose that the optimal solution to \eqref{Pc_PG_diag} denoted by $\hat{\mathbf{\Lambda}}_z^\star$ is non-diagonal. Then let $\breve{\mathbf{\Lambda}}_z^\star \triangleq \diag\{(\hat{\mathbf{\Lambda}}_z^\star)_{1,1},\ldots,(\hat{\mathbf{\Lambda}}_z^\star)_{M,M}\}$. One can easily see that $\breve{\mathbf{\Lambda}}_z^\star$ has a smaller objective than $\hat{\mathbf{\Lambda}}_z^\star$ in \eqref{Pc_PG_diag}, which contradicts with the optimality of $\hat{\mathbf{\Lambda}}_z^\star$. Hence $\hat{\mathbf{\Lambda}}_z^\star$ must be diagonal.
Let $\hat{\mathbf{\Lambda}}_z=\diag\{\bm{\eta}\}$, and then problem \eqref{Pc_PG_diag} can be equivalently changed to the simplex projection problem in \eqref{Pc_PG_simplex}. From $\diag\{\bm{\eta}^\star\}=\mathbf{U}^H\mathbf{\Lambda}_z^\star\mathbf{U}$, we obtain the structure of the optimal solution as shown in \eqref{Pc_PG_sol}.

As for the simplex projection problem in \eqref{Pc_PG_simplex}, it is well studied in \cite{Palomar2005a} and its optimal solution is given by the water-filling solution in \eqref{Pc_PG_sol_eta}. This completes the proof.
\end{appendices}

%\bibliographystyle{IEEEtran}
%\bibliography{my}

\end{document}